\documentclass[12pt]{article}
\usepackage{amsfonts}

\input epsf.tex
\title{Conformal Field Theory of the \\
Integer Quantum Hall Plateau Transition} 
\author{Martin R. Zirnbauer\footnote{Current address:
  Department of Mathematics, Fine Hall, Princeton University, USA.
  Permanent address: Institut f\"ur Theoretische Physik, Universit\"at
  zu K\"oln, Z\"ulpicher Str. 77, 50937 K\"oln, Germany.  
  Email: zirn@thp.Uni-Koeln.DE}}
\begin{document}
\date{May 28, 1999}
\maketitle
\begin{abstract}
A solution to the long-standing problem of identifying the conformal
field theory governing the transition between quantized Hall plateaus
of a disordered noninteracting 2d electron gas, is proposed.  The
theory is a nonlinear sigma model with a Wess-Zumino-Novikov-Witten
term, and fields taking values in a Riemannian symmetric superspace
based on ${\rm H}^3 \times {\rm S}^3$.  Essentially the same conformal
field theory appeared in very recent work on string propagation in
${\rm AdS}_3$ backgrounds.  We explain how the proposed theory manages
to obey a number of tight constraints, two of which are constancy of
the partition function and noncriticality of the local density of
states.  An unexpected feature is the existence of a truly marginal
deformation, restricting the extent to which universality can hold in
critical quantum Hall systems.  The marginal coupling is fixed by
matching the short-distance singularity of the conductance between two
interior contacts to the classical conductivity $\sigma_{xx} = 1/2$ of
the Chalker-Coddington network model.  For this value, perturbation
theory predicts a critical exponent $X_{\rm t}=2/\pi$ for the typical
point-contact conductance, in agreement with numerical simulations.
The irrational exponent is tolerated by the fact that the symmetry
algebra of the field theory is Virasoro but not affine Lie algebraic.
\end{abstract}

\section{Introduction}

A two-dimensional electron gas subjected to strong magnetic fields
exhibits the striking phenomenon of the quantum Hall effect
\cite{klitzing,tsg}: as the temperature is lowered into the
(sub-)Kelvin range, the Hall conductance evolves towards a staircase
function, with quantized plateau values $\nu e^2/h$ occurring around
certain magic Landau level filling fractions $\nu$.  It is fair to say
that the physical reasons for the quantization and its stability with
respect to changing the magnetic field, are by now well understood
\cite{prange,hajdu}.  What has proved more difficult to clarify is the
precise nature of the {\it transitions between plateaus}.  For all we
know, these are associated with quantum critical points of the
electron gas.  Extrapolation to zero temperature and infinite system
size indicates a second-order phase transition with a divergent
correlation length.  In the integer version of the effect, which we
will be concerned with in the present paper, interactions between the
electrons are thought to be irrelevant, and the critical behavior is
attributed to the interplay between the disorder, which tends to
localize the electrons, and the kinetic energy quenched by the strong
magnetic field, causing delocalization for some isolated values of the
Fermi energy \cite{huckestein}.  The major theoretical goal is to
identify the proper {\it low-energy effective field theory}.  On
general grounds, this theory is expected to be conformal at the
transition, and should provide a systematic framework for the
calculation of critical exponents and the formulation of a complete
scaling theory \cite{wei,pruiscal}.  Needless to say, in spite of
fifteen years of massive effort, that ambitious theoretical goal was
never achieved.

The early development of the subject was guided by a field theory of
the type of a $G / H$ nonlinear sigma model, called Pruisken's model
\cite{pruisken}.  Its original formulation relied on a fermionic 
version of the replica trick, leading to a symmetry group $G = {\rm
U}(2n)$ (with $n = 0$) gauged on the right by $H = {\rm U}(n) \times
{\rm U}(n)$.  A mathematically satisfactory formulation avoiding
replicas was introduced by Weidenm\"uller \cite{haw}, who employed the
supersymmetric method of Efetov \cite{efetov}.  In that variant, $G$
is a pseudo-unitary supergroup ${\rm U}(1,1|2)$, and $H = {\rm U}(1|1)
\times {\rm U} (1|1)$.  In either case, the field theory has two 
dimensionless coupling constants, denoted by $\sigma_{xx}$ and
$\sigma_{xy}$, which are identified with the longitudinal and Hall
conductivities of the disordered electron gas.  On the basis of
Pruisken's model, Khmelnitskii \cite{khmelnitskii} conjectured a
renormalization group flow diagram for $\sigma_{xx}$ and
$\sigma_{xy}$, where the central feature is a fixed point of the flow
at some $\sigma_{xx}^*$ and $\sigma_{xy}^* = 1/2$.  The fixed point
was argued to have one relevant ($\sigma_{xy} - 1/2$) and one
irrelevant ($\sigma_{xx} - \sigma_{xx}^*$) perturbation, giving a fair
representation of experimental and numerical data on the transition.
Unfortunately, the initial successes and guesses inspired by
Pruisken's nonlinear sigma model were not backed up by a more complete
solution.  The problem is that the postulated fixed point --- if it
exists as a fixed point of Pruisken's model, which is not really clear
--- lies at strong coupling, or small $\sigma_{xx}$, where it cannot
be controlled by its representation through $G /H$ fields.  (The
one-instanton computations of Pruisken and collaborators \cite{llp}
can only be trusted at weak coupling, and to learn something about the
fixed point, a far extrapolation to strong coupling is required.)
Consequently, no quantitative results beyond the rough, though
instructive, picture of two-parameter scaling have ever come out of
that theory.

The disillusionment about Pruisken's model grew into a forceful
complaint \cite{afflecki} when a detailed understanding of a related
2d fixed point became available, namely that governing the six-vertex
model with isotropic vertex weights or, equivalently, the sine-Gordon
model at $\beta^2 = 8\pi$, or, the $xy$-model at the
Kosterlitz-Thouless transition.  Another of its many incarnations is
found in isotropic 1d quantum antiferromagnets with half-integral
spin, which following Haldane \cite{haldane} map on the ${\rm O}(3)$
nonlinear sigma model with a topological term and topological angle
$\theta = \pi$.  In that paradigmatic system, it soon became clear
\cite{ah} that the ``good'' theory, where conformal invariance is
manifest and critical properties can be computed in great detail, is
not the ${\rm O}(3)$ model but a Wess-Zumino-Novikov-Witten (WZW)
model.  Characteristically, the fields of the latter take values in a
group (which is ${\rm SU}(2)$ here), whereas the target space of the
${\rm O}(3)$ model is a coset space ${\rm SU}(2)/{\rm U}(1)$.  Thus
the field is promoted from being coset-valued to group-valued, at
criticality.  Although antiferromagnets seem to be a far cry from
disordered electrons, the lesson learned from this example does have a
bearing on the quantum Hall critical point and ought to be taken
seriously: Pruisken's replicated model at replica number $n = 1$ (and
critical $\sigma_{xy} = 1/2$) is nothing else than the ${\rm O}(3)$
model at $\theta = \pi$, which in turn is also a basic constituent of
the supersymmetric $G/H$ model.

Drawing on the insight gained from antiferromagnets, a natural idea
for making progress with the quantum Hall plateau transition is to try
and promote the field space $G/H$ to a group or group-like manifold.
(Note that we are well advised to avoid the mathematical ambiguities
\cite{vz,replica} of the replica trick and work with the
supersymmetric formulation.  Affleck attempted to execute the program
in the replicated version, but failed \cite{affleckf}.)  Although this
idea has now been around for more than a decade, progress was
painfully slow.  The reason was that there existed a number of severe
obstacles, a chronological personal account of which is as follows.

Given the symmetries of the supersymmetric version of Pruisken's
model, one might take a WZW model with target ${\rm U}(1,1)|2)$, or
${\rm SU}(1,1|2)$, or ${\rm PSU}(1,1|2)$, for a first candidate.
Here we stall immediately.  In trying to solve the statistical physics
problem at hand, we have to be very discriminating about which
functional integral to accept as well-defined and which not.  In
concrete terms, we are looking for a field theory defined over {\it
Euclidian} two-space, and with a target space of {\it Euclidian}
signature.  This constraint eliminates candidate theories with an
action functional that is bounded neither from below nor from above.
Among these are the above supergroups, the natural supergeometry of
which is non-Riemann, or of indefinite signature.  (The natural
geometry is forced on us by symmetry considerations.)

Let us mention in passing that, to have a WZW model with definite
metric, one option is to start from $G = {\rm U}(1,1|2)$ and gauge by
$H = {\rm U}(1|1) \times {\rm U}(1|1)$.  What we have in mind here is
the functional integral version \cite{gk} of the Goddard-Kent-Olive
construction.  Unfortunately, in that construction the gauge group
acts by {\it conjugation} (so it acts simultaneously on the left and
right), which ruins conservation of some of the $G$ currents.
Symmetries are not lost under renormalization, and because $G$ is the
symmetry group of (the supersymmetric version of) Pruisken's model, it
{\it must} be present in the fixed-point theory.  This kills the idea
of gauging the $G$-WZW model by $H$.

The next attempt is to modify ${\rm U}(1,1|2)$, so as to arrive at a
target manifold with better metric properties, giving an action
functional bounded from below by the constant fields.  Curiously, it
turns out that the proper modification does not exist within the realm
of standard supermanifold theory, but requires the introduction of
objects I call {\it Riemannian symmetric superspaces} \cite{suprev}.
They belong to the general category of cs-manifolds \cite{bernstein}.
The definition of these nonstandard notions, and their illustration at
a well-chosen simple example, will be given in Section
\ref{sec:motiv}.  The crucial feature of the ``good'' variant of ${\rm
U}(1,1|2)$ --- let's name it {\bf X} for short --- is that it is based
on a symmetric space $M_{\rm B} \times M_{\rm F}$, $M_{\rm B} = {\rm
GL}(2,{\mathbb C}) / {\rm U}(2)$ and $M_{\rm F} = {\rm U}(2)$, which
has the desired property of being {\it Riemann} in the geometry
inherited from the natural supergeometry of ${\bf X}$. In some sense,
${\rm U}(1,1|2)$ is one ``real'' form of the complex supergroup ${\rm
GL}(2|2)$, and ${\bf X}$ is another.  (In the notation of Section
\ref{sec:motiv}, ${\rm X}$ is the symmetric superspace $\left( {\rm
GL}(2|2) , {\rm STr} ( g^{-1} {\rm d}g)^2 , M_{\rm B} \times M_{\rm F}
\right)$.)  While ${\bf X}$ is {\it not} a group, it does give rise to
an acceptable and well-defined functional integral, the ${\bf X}$-WZW
model.  (By an abuse of terminology, one might also call it the ${\rm
GL}(2|2)$ WZW model \cite{rs}.)

Aside from its distinguished mathematical role, the ${\bf X}$-WZW
model also has a noteworthy {\it physical} origin.  Several models
have been proposed as a starting point for the description of the
quantum Hall plateau transition, and among them there is a model
\cite{lfsg} of two-dimensional Dirac fermions subject to various 
types of disorder: random vector potential, random scalar potential,
and random mass.  For the purpose of computing the disorder-averaged
Green functions and other quenched correlation functions of the
model, one adds a bosonic $\beta$-$\gamma$ ghost system, which
normalizes the partition function to unity \cite{denis}.  By a
superextension of Witten's nonabelian bosonization scheme
\cite{nonab}, the weakly disordered Dirac-$\beta\gamma$ system then
transforms into a weakly perturbed WZW model.  Actually, there exist
two schemes \cite{affleckprl}, and the better one to use in the
present context leads to the target being ${\rm GL}(1|1)$ for a single
Green function, and ${\rm GL}(2|2)$ for two Green functions.
These identifications of the target space are naive, and a careful
analysis of the second case shows that proper use of the bosonization
scheme leads to the target ${\bf X}$.  Thus the pure
Dirac-$\beta\gamma$ system is equivalent to the ${\bf X}$-WZW model
(at level $k = 1$), and the disorder which is present translates into
perturbations of it.

The usefulness of such an approach now hinges on the nature of the
perturbations.  From \cite{lfsg} it is known that the random scalar
potential, the random vector potential and the random mass are
marginally relevant, truly marginal, and marginally irrelevant, in
that order.  The dangerous perturbation by a random scalar potential,
which is generic to the quantum Hall (QH) universality class, grows
under renormalization and drives the Dirac-$\beta\gamma$ system to an
unknown fixed point at strong coupling.  Unfortunately, nonabelian
bosonization does not give much of a clue as to how to handle that
strong-coupling problem.  Presumably, the fixed point can somehow be
described in terms of fields taking values inside ${\bf X}$, but
exactly what happens remains mysterious.  (The situation is less
favorable here than in Affleck's trick \cite{affleckprl} for passing
from the critical Heisenberg chain via Dirac fermions to the ${\rm
SU}(2)_1$ WZW model.  In that case, the relevant perturbation simply
reduces ${\rm U}(2)$ to ${\rm SU}(2)$.)  Thus we are stuck once again.

The next piece of confusing evidence came from a 1996 numerical study
\cite{jmz} of point-contact conductances in the Chalker-Coddington
network model of the QH plateau transition.  From the perspective of
conformal field theory, the point-contact conductance is the most
basic and ``clean'' observable, as Pruisken's model expresses it as a
{\it two-point function of local fields}.  In a WZW model, the scaling
dimensions of local (primary) fields $\phi_\lambda$ transforming
according to a representation $\lambda$, are given \cite{kz} by the
formula ${\cal C}_\lambda / (k + h_*)$, where ${\cal C}_\lambda$ is
the quadratic Casimir invariant evaluated on $\lambda$, and the level
$k$ and the dual Coxeter number $h_*$ are integers.  In view of this
fact, it has to be termed striking that the critical exponent $X_{\rm
t}$ for the typical point-contact conductance was found be $X_{\rm t}
= 0.640 \pm 0.009$, which is numerically close to $2/\pi \approx
0.637$.  The latter value, if exact, is hard to reconcile with the
above formula, which predicts rational numbers.  It seemed, at that
point, that the idea of promoting the field space $G/H$ to a group (or
group-like manifold), and passing from Pruisken's model to a WZW
model, fails to work.

The present paper was triggered by the recent appearance of two
articles \cite{bvw,bzv} related to superstring propagation on ${\rm
AdS}_3$ backgrounds.  There, a prominent role is played by a nonlinear
sigma model with target ${\rm PSU}(1,1|2)$, the supergroup obtained
from ${\rm U}(1,1|2)$ by requiring unit superdeterminant and gauging
w.r.t.~the multiples of the unit matrix.  The intriguing message from
those articles is that the ${\rm PSU}(1,1|2)$ nonlinear sigma model is
{\it conformal at any value of its coupling}, $f$.  The model also
allows for the presence of a Wess-Zumino term with topological
coupling $k$, so we have a two-parameter family of conformal field
theories at our disposal.  (In the string-theory context, the two
coupling constants are related to the Ramond-Ramond and Neveu-Schwarz
fluxes that are due to a number of fivebranes wrapped around some
Calabi-Yau manifold.)  The marginality of the coupling $f$ looks very
promising from our perspective, as it suggests enough flexibility to
accommodate the peculiar critical exponent found for the typical
point-contact conductance.  We are thus led to reconsider the WZW
model idea.

Our tale does not converge to a quick conclusion, as there still
exist a number of difficulties to overcome.  First of all, ${\rm
PSU}(1,1|2 )$ is one of those target spaces we discarded right at the
outset, on the grounds that we insist on having a target metric with
Euclidian signature.  Second, the Virasoro central charge of the ${\rm
PSU}(1,1 |2)$ nonlinear sigma model (with and without Wess-Zumino
term) has the value $c = -2$, which is at variance with a basic
constraint on the theory: its partition function must be identically
equal to unity, independent of all parameters of the noninteracting
electron gas.  Third, correlation functions that involve only retarded
or only advanced Green functions of the disordered electron system,
are known to be noncritical (or even trivial).  It is not a priori
obvious how one can arrange for the ${\rm PSU}(1,1|2)$ model to
reproduce this feature.  Fourth, the marginality of the coupling $f$,
while needed to accommodate the exponent $X_{\rm t}$, appears to be at
odds with the observation of universal critical behavior in QH
systems.

In the present paper, these problems will be addressed and solved.  In
brief, the first one (indefinite metric) is overcome by trading ${\rm
PSU}(1,1|2)$ for a submanifold of the Riemannian symmetric superspace
${\bf X}$, which is obtained by dividing out ${\mathbb R}^+ \subset
M_{\rm B}$ and ${\rm U}(1) \subset M_{\rm F}$.  The second one
(central charge $c \not= 0$) is rectified by postulating the existence
of a bosonic ghost field that can alternatively be regarded as forming
part of the functional integration measure.  The third one
(noncriticality of all correlation functions that probe only the
retarded or advanced sector) turns out to be resolved as a consequence
of BRST invariance of the supersymmetric theory with properly defined
target.  Concerning the fourth point (marginal coupling $f$), we shall
argue that universality prevails to the extent that the conductivity
governing the {\it classical} or incoherent transport near absorbing
boundaries is universal.

Let us now summarize the plan of the paper.  We start out by reviewing
in Sections \ref{sec:cc}--\ref{sec:spin} three cornerstones of the
theory of the QH plateau transition: the network model of Chalker and
Coddington, the supersymmetric version of Pruisken's nonlinear sigma
model, and an antiferromagnetic superspin chain.  (Along the way, we
point out an exact mapping from the autocorrelation function of
spectral determinants of the network model, to a perturbed six-vertex
model.) The physical information drawn from them is condensed into a
check list of conditions the fixed-point theory must satisfy, in
Section \ref{sec:constraints}.  We then change gears and elucidate, in
Section \ref{sec:motiv}, the notions of Riemannian symmetric
superspace and cs-manifold, which are needed for the definition of the
target space and its invariant Berezin integral in Sections
\ref{sec:target} and \ref{sec:berezin}.  A candidate for the fixed
point is proposed in Section \ref{sec:cft}.  Normalization of its
partition function and triviality of the BRST invariant correlation
functions is demonstrated in Section \ref{sec:brst}.  Section
\ref{sec:conformal} briefly reviews the arguments for conformal
invariance of the model, and Section \ref{sec:checks} checks more
items of our list.  The marginal coupling $f$ is fixed in Section
\ref{sec:coupling}, by matching the short-distance singularity of the
conductance between two interior contacts to the classical
expectation.  With the value of $f$ thus determined, we argue in
Section \ref{sec:quantum} that the algebraic decay of the typical
point-contact conductance for the network model is governed by the
irrational exponent $X_{\rm t} = 2/\pi$, in agreement with the
numerics.  An assessment of where the theory now stands and where it
will go, is given in the last section.

Finally, a word of warning is in order.  This paper addresses an
audience including disordered electron physicists, conformal field
theorists, and high-energy physicists.  Therefore, an effort was made
to explain some trivial things.  On the other hand, part of the
material, particularly in the later sections, is too complex to be
treatable below a certain miminum of mathematical sophistication, and
some basic familiarity with the theory of symmetric spaces,
supermanifolds, and harmonic analysis had to be assumed.

\section{Network model}
\label{sec:cc}

The purpose of the present paper is to propose a field-theory
Lagrangian describing the critical behavior at the transition between
two neighboring plateaus of the integer quantum Hall effect.  Although
our proposal is not constructive (in the sense of providing a complete
sequence of steps leading from a microscopic model to the field
theory), it will take a number of important clues and constraints from
the representation of the quantum Hall universality class by
noninteracting disordered electrons in a strong magnetic field.  A
particularly neat and efficient representative of this universality
class is the network model of Chalker and Coddington \cite{cc,lwk},
which we are now going to review.

In its original formulation, the model was conceived as a device for
computing the transfer of electron wave amplitudes across a finite
two-dimensional quantum Hall sample.  A wave function of the model is
defined to be a set of complex amplitudes, one for each edge or link
of a square network.  A characteristic feature, originating from the
presence of a strong magnetic field, is the unidirectional motion
specified by arrows, see Figure 1.  The elementary building blocks of
the model are $2 \times 2$ scattering matrices $S$ assigned to the
vertices or nodes of the network.  Being elements of the unitary group
${\rm U}(2)$, these matrices can be written as
	$$ 
	S = \pmatrix{ 	{\rm e}^{i\varphi(o_1)} &0\cr 
			0 &{\rm e}^{i\varphi(o_2)} \cr} 
	\pmatrix{\cos\rho &\sin\rho\cr -\sin\rho &\cos\rho\cr} 
	\pmatrix{	{\rm e}^{i\varphi(i_1)} &0\cr 
			0 &{\rm e}^{i\varphi(i_2)} \cr} \;.  
	$$ 
Each phase factor ${\rm e}^{i\varphi}$ belongs to one link of the
network.  Disorder is introduced by taking the phase factors to be
independent identically distributed random variables drawn from ${\rm
U}(1)$.  Averages over the disorder will be denoted by $\langle
... \rangle$.  The distribution of the phase factors is taken to be
uniform on ${\rm U}(1)$, which results in the model having a local
${\rm U}(1)$ lattice gauge invariance.  The parameter $\rho$ is taken
to be fixed (as opposed to random) and homogeneous over the network,
and determines the probability for scattering to the left or right at
each node to be $p_{\rm L}=|\cos\rho|^2$ or $p_{\rm R}=|\sin\rho|^2$.
The connection rules specified by the scattering matrices on the nodes
define a transfer matrix for the total system.  The network model is
critical when the probabilities for scattering to the left and right
are equal: $p_{\rm L} = p_{\rm R} = 1/2$.  (We mention in passing that 
the approximations leading to the model are readily justified for 
slowly varying random potentials.)
\begin{figure}
	\hspace{4.5cm} 
	\epsfxsize=4cm \epsfbox{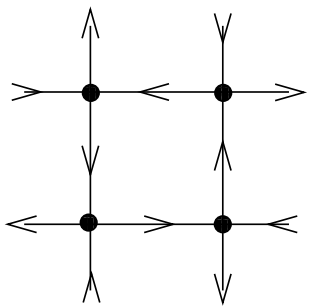}
        \caption{Unit cell of the network model.}
        \label{fig:network}
      \end{figure}

Although the transfer matrix picture has its merits, it will be more
useful for our purposes to think about the network model in another
way, namely as a {\it dynamical} system with discrete time \cite{km1}.
The evolution operator for one time step is a unitary operator denoted
by $U$, and the dynamics is generated by iterating $U$.  Thus, if the
state of the electron at time $t = 0$ is $|\psi_0\rangle$, the state
after $n$ elementary time steps is $|\psi_{t = n}\rangle = U^n|\psi_0 
\rangle$.  The time evolution operator $U$ is a product of two factors:
$U = U_0 U_1$.  The first of these ($U_1)$ encodes the deterministic
part of the scattering at the nodes.  The other factor $(U_0)$
multiplies the wave function on each link $l$ by the corresponding
random ${\rm U}(1)$ element ${\rm e}^{i\varphi(l)}$.

When the network model is viewed as a quantum dynamical system, we can
define for it an analog of the resolvent operator $(E \pm i\epsilon -
H)^{-1}$ of a Hamiltonian system with energy $E$ and Hamiltonian $H$.
Recalling the identity
	$$
	(E + i\epsilon - H)^{-1} = - i \int_0^\infty 
	{\rm e}^{it(E + i\epsilon - H)} dt 
	\qquad (\epsilon > 0) \;,
	$$
we do the following.  We replace $E+i\epsilon$ by a complex
quasi-energy $i^{-1}\ln z$ with $|z| < 1$, the time evolution operator
${\rm e}^ {-itH}$ by $U^n$, and the integral over continuous time $t$
by a sum over discrete time $n$.  The above identity then transcribes
to
	$$
	(1 - zU)^{-1} = \sum_{n=0}^\infty z^n U^n \;.
	$$
Thus the correct analog of the operator $(E + i\epsilon-H)^{-1}$ is
$(1 - z U)^{-1}$, and the analog of $(E-i\epsilon-H)^{-1}$ is $(1 -
\bar z \bar U)^{-1}$.  Many observables of the network -- as an
example we mention the point-contact conductance \cite{jmz} -- can
be expressed as products of matrix elements of these operators.

Although its primary realm of application is charge transport, the
network model has also been profitably used for studying spectral
correlations \cite{km2}.  In that case one takes the network to be
{\it closed}, and views the eigenvalues of $i \ln U$ as quasi-energy
``levels'' on the interval $[0,2\pi]$.  One quantity of interest is 
the so-called two-level correlation function $R_2(\omega)$, which has
a well-known expression as the (discrete) Fourier transform of the 
quantum return probability $\langle |{\rm Tr} \, U^n |^2 \rangle$:
	$$
	R_2(\omega) = (2\pi^2)^{-1} \sum_{n=1}^\infty \cos(n\omega) 
	\left\langle \big| {\rm Tr} \, U^n \big|^2 \right\rangle \;.
	$$
Another quantity that has been the focus of recent work \cite{kt} is the 
autocorrelation function of spectral determinants,
	$$
	C(\omega) = \left\langle {\rm Det}(1 - {\rm e}^{i\omega/2} U)
	\, {\rm Det}(1 - {\rm e}^{i\omega/2} \bar U) \right\rangle \;.
	$$
More generally, we can consider the following correlation function:
	\begin{equation}
	\Omega(a_0,a_1;b_0,b_1) = \left\langle
	{{\rm Det}(1 - a_1 U) \, {\rm Det}(1 - b_1 \bar U) \over
	{\rm Det}(1 - a_0 U) \, {\rm Det}(1 - b_0 \bar U)}
	\right\rangle \;.
	\label{speccorr}	
	\end{equation}
We recover $C(\omega)$ from it by setting $a_0 = b_0 = 0$ and $a_1 =
b_1 = {\rm e}^{i\omega/2}$.  We can also extract $R_2(\omega)$, by
differentiating with respect to $a_0$ and $b_0$ at $a_0 = a_1 = b_0 =
b_1 = {\rm e}^{i\omega/2}$ and then taking the real part.  A
(perturbed) conformal field theory formulation of the correlator
$\Omega$ at criticality will be proposed in Section \ref{sec:cft}.  
Here we wish to point out the following two features.  Firstly, when
$a_0$ is set equal to $a_1$, or $b_0$ equal to $b_1$, the correlator
becomes trivial:
	$$
	\Omega(a,a;b_0,b_1) = 1 = \Omega(a_0,a_1;b,b) \;.
	$$
Indeed, in the former case the first determinants in the numerator and
denominator cancel, leaving $\langle {\rm Det}(1-b_1 \bar U) / {\rm
Det}(1 - b_0 \bar U) \rangle$.  For $|b_0| < 1$ this can be expanded 
in a convergent power series in $\bar U$.  Since $\langle {\rm e}^{-in
\varphi}\rangle=0$ for $n>0$, only the very first term  in the series 
survives disorder averaging, giving the trivial result of unity.  The
same argument applies to the latter case ($b_0 = b_1 = b$).  For
similar reasons, any correlation function or observable involving only
retarded information $(U^n)$ or only advanced information $(\bar U^n)$
is trivial.  This puts a strong constraint on any field theory that is
to be a serious candidate for the QH plateau transition.

Secondly, recall that setting $a_0 = b_0 = 0$ and $a_1 = b_1 = {\rm
e}^{i\omega/2}$ yields the correlator $C(\omega)$.  Its critical
behavior has recently been argued \cite{kt} to fall in a much studied
and well understood universality class that has an ${\rm SU}(2)$
invariance.  This is the universality class of the six-vertex model
with isotropic vertex weights, or the one-dimensional isotropic
spin-1/2 Heisenberg antiferromagnet, or the sine-Gordon model at
$\beta^2 = 8\pi$, or the $xy$-model at the Kosterlitz-Thouless
temperature.  The conformal field theory governing this class is known
\cite{affleckprl} to be the Wess-Zumino-Novikov-Witten (WZW) model of
the group ${\rm SU}(2)$ at level $k = 1$.  Thus another condition to
impose on the theory we are looking for, is that it has to flow to
the ${\rm SU}(2)_1$ WZW model on sending $\ln a_0$ and $\ln b_0$ to
minus infinity.

While the derivation given in \cite{kt} took the thorny route of the
${\rm O}(3)$ nonlinear sigma model, the critical nature of the
correlator $C(\omega)$ can be understood quite directly and
convincingly from the network model.  Let us take a break from the
general development and provide a few details here.  We start off by
representing the product of spectral determinants as a Gaussian
integral over (retarded and advanced) anticommuting fields
$\psi_\uparrow(l)$ and $\psi_\downarrow(l)$ placed on the links $l$ of
the network.  The average over the random phase factors ${\rm
e}^{i\varphi(l)}$ is then carried out by using at every link the
identity
	$$
	\int_0^{2\pi} d\varphi \, \exp \left( 
	{\rm e}^{i\varphi} \bar\psi_\uparrow\psi_\uparrow 
	+ {\rm e}^{-i\varphi} \bar\psi_\downarrow\psi_\downarrow \right)
	= 2 \int_{\mathbb C} {dz d\bar z \over (1+\bar zz)^3} \,
	\exp \left( z\bar\psi_{\uparrow}\psi_\downarrow - 
	\bar z\bar\psi_\downarrow\psi_\uparrow \right) \;,
	$$
which is a special case of the ``color-flavor transformation''
\cite{icmp97} and is elementary to verify by Taylor expansion of the
integrand on both sides.  A beautiful feature of the transformation is
that it preserves the Gaussian dependence on $\psi$.  We can therefore
integrate out $\psi$ again, and arrive at another determinant, now 
with matrix entries that depend on the complex field $z(l)$.  By the
structure of the network model, the determinant factorizes as a 
{\it product} of terms, or weights, one for each vertex of the 
network.  At the critical point $p_{\rm L} = p_{\rm R} = 1/2$, and 
for a vertex with incoming links $i_1$ and $i_2$, and outgoing links 
$o_1$ and $o_2$, the weight is \cite{network}
	$$
	{1 + {1\over 2}{\rm e}^{i\omega} \left( \bar z(o_1) + \bar
	z(o_2) \right) \left( z(i_1) + z(i_2) \right) + {\rm e}^{2i\omega}
	\bar z(o_1) \bar z(o_2) z(i_1) z(i_2) \over 
        \sqrt{1+\bar z(o_1)z(o_1)} \sqrt{1+\bar z(o_2)z(o_2)}
        \sqrt{1+\bar z(i_1)z(i_1)} \sqrt{1+\bar z(i_2)z(i_2)} }\;.
	$$
Hence the transformed theory has the structure of a {\it vertex model}.
To bring it into a familiar form, one has to recognize the intrinsic
meaning of the complex number $z(l)$ (for a fixed link $l$) as a
parameter for coherent spin-1/2 states: $|\uparrow \rangle + z |
\downarrow \rangle$.  (This meaning is particularly evident from the
general proof \cite{circular} of the color-flavor transformation.)  By
interpreting the vertex weights as matrix elements between coherent
states, and using the closure relation
	$$
	{1\over\pi} \int_{\mathbb C} {dz d\bar z \over (1+\bar zz)^3} \,
	\Big( |\uparrow\rangle + z |\downarrow\rangle \Big)
	\Big( \langle\uparrow| + \bar z\langle\downarrow| \Big)
	= |\uparrow \rangle \langle \uparrow| + |\downarrow \rangle
	\langle \downarrow| \;,
	$$
one can rewrite the correlator $C(0)$ as the partition sum of a vertex
model for classical variables taking two values, $\uparrow$ and
$\downarrow$.  The vertex weights can be extracted from the above
expression by observing that $z = 0$ means spin up, and $z = \infty$
spin down. For $\omega = 0$ the vertex weights are isotropic, or ${\rm
SU}(2)$-symmetric, giving a model that belongs to a family known as
the six-vertex model in the area of integrable systems \cite{baxter}.
Note that the calculation we have sketched is free of approximations,
so the autocorrelation function of spectral determinants $C(0)$ at the
critical point is {\it exactly equal} to the partition sum of the
six-vertex model at its ${\rm SU}(2)$-symmetric point.  The parameter
$\omega$ acts as a perturbation breaking ${\rm SU}(2)$ symmetry as
well as criticality.  (In the anisotropic limit of the six-vertex
model as an $xxx$ quantum spin chain with Hamiltonian $H$, the system
is perturbed by coupling to a staggered imaginary field, $H \to H +
i\omega \sum_n (-1)^n S_n^z$.  For a real field, this Hamiltonian has
been studied by Oshikawa and Affleck \cite{oa}, and by Essler and
Tsvelik \cite{et}, using abelian bosonization.)

While the equivalence between the network model correlator $C(\omega)$
and a (perturbed) six-vertex model is interesting in its own right and
deserves further study, we will not pursue it here, as it does not
form the subject of the present paper.  We have pointed out the
equivalence because it adds much support to the claim \cite{kt} that
one basic building block of the field theory we are seeking is the
conformal field theory limit of the six-vertex model, namely the ${\rm
SU}(2)_1$ WZW model.

\section{Pruisken's nonlinear sigma model}
\label{sec:nlsm}

Historically, the network model of Chalker and Coddington was preceded
by Pruisken's nonlinear sigma model, which dominated the early efforts
to understand electron delocalization in the integer quantum Hall
effect.  This important cornerstone of the theory will be reviewed next.

The task of computing transport coefficients for a disordered electron
system amounts to calculating disorder averages of products of
retarded and advanced electron Green functions.  A good way of going
about it is to map the problem on an effective field theory.  The
initial development \cite{wegner1,sw,elk} of the subject relied on the
replica trick for computing disorder averages.  For the case of
systems in a {\it weak} magnetic field, this led to a nonlinear sigma
model with target space $G/H$, where $G/H={\rm U}(n,n)/{\rm
U}(n)\times{\rm U}(n)$ or ${\rm U}(2n)/{\rm U} (n)\times{\rm U}(n)$
with $n=0$ for bosonic resp.~fermionic replicas.  A mathematically
satisfactory variant based on the supersymmetric formalism emerged
from the pioneering work of Efetov \cite{efetov}.  In that
formulation, $G = {\rm U}(1,1|2)$, the group of pseudo-unitary
$4\times 4$ supermatrices $g$ preserving an indefinite Hermitian form:
	$$
	g^\dagger \eta g = \eta \;, \qquad 
	\eta = {\rm diag}(-1,+1,+1,+1) \;,
	$$ 
and $H = {\rm U}(1|1) \times {\rm U}(1|1)$.  (For many purposes, it is
preferable to work with the complexified group $G_{\mathbb C} = {\rm
GL}(2|2)$ but we will stick with $G$ for now.)  The Lagrangian of the
theory is conventionally presented in terms of a field denoted by $Q$:
	$$ 
	Q = g \, \Sigma_3 \, g^{-1} \;, \qquad 
	\Sigma_3 = {\rm diag}(+1,+1,-1,-1) \;.  
	$$ 
The diagonal matrix $\Sigma_3$ discriminates between the retarded $(+)$ 
and advanced $(-)$ sector of the theory.  A natural parametrization of 
the nonlinear field $Q$ is
	$$ 
	Q = \pmatrix{1 &Z\cr \tilde Z &1\cr} \pmatrix{1 &0\cr
	0 &-1\cr} \pmatrix{1 &Z\cr \tilde Z &1\cr}^{-1} \;, 
	$$ 
where 
	\begin{equation}
	Z = \pmatrix{Z_{\rm BB} &Z_{\rm BF}\cr Z_{\rm FB} &Z_{\rm FF}\cr} \;,
	\qquad \tilde Z = \pmatrix{\tilde Z_{\rm BB} &\tilde Z_{\rm BF}\cr
	\tilde Z_{\rm FB} &\tilde Z_{\rm FF}\cr} 
	\label{ztz}
	\end{equation}
are complex $2 \times 2$ supermatrices, with $\tilde Z_{\rm FF} = - 
\bar Z_{\rm FF}$ and $\tilde Z_{\rm BB} = + \bar Z_{\rm BB}$.  (We do 
not specify how complex conjugation relates $Z_{\rm BF}, Z_{\rm FB}$ to
$\tilde Z_{\rm FB},\tilde Z_{\rm BF}$, as there is no need for that.)
The variable $Z_{\rm FF}$ takes values in ${\mathbb C}$ with a point
added at infinity, which is the same as a two-sphere ${\rm S}^2
\simeq {\rm U}(2)/{\rm U}(1) \times {\rm U}(1)$.  The range of $Z_{\rm
BB}$ is restricted by $|Z_{\rm BB}|^2 < 1$.  This is Poincar\'e's
model of the two-hyperboloid ${\rm H}^2 \simeq {\rm U}(1,1)/{\rm U}(1)
\times {\rm U}(1)$.  The Lagrangian of the field theory is
	\begin{eqnarray*}
	L_0 &=& - {\sigma_{xx} \over 8} \, {\rm STr} \, \partial_\mu Q
	\, \partial_\mu Q \\
	&=& \sigma_{xx} \, {\rm STr} \, (1-\tilde ZZ)^{-1} 
	\partial_\mu \tilde Z (1-Z\tilde Z)^{-1} \partial_\mu Z \;,
	\end{eqnarray*}
where ${\rm STr}$ means the supertrace.  The coupling constant
$\sigma_{xx}$ has an interpretation as the dissipative conductivity of
the electron gas (conductances being measured in natural units
$e^2/h$).  In two space dimensions, the coupling $\sigma_{xx}$ is
dimensionless, and from renormalization-group assisted perturbation
theory one expects the existence of a mass gap, implying that all
electron states are localized in that case.

Pruisken's insightful contribution \cite{pruisken} was to add to the
Lagrangian a topological density, $L_{\rm top}$.  Such a term exists,
and is nontrivial, both for the fermionic replica theory and for the
supersymmetric theory.  The latter version was first formulated by
Weidenm\"uller \cite{haw}, and reads
	\begin{eqnarray*}
	L_{\rm top} &=& {\sigma_{xy} \over 8} \, \epsilon_{\mu\nu}
	\, {\rm STr} \, Q \, \partial_\mu Q \, \partial_\nu Q \\
	&=& \sigma_{xy} \, \epsilon_{\mu\nu} \, {\rm STr} \, 
	(1-\tilde ZZ)^{-1} \partial_\mu \tilde Z 
	(1-Z\tilde Z)^{-1} \partial_\nu Z \;.
	\end{eqnarray*}
The coupling constant $\sigma_{xy}$ is identified with the Hall
conductivity of the two-dimensional electron gas.  Pruisken's key idea
was that such a term is needed to break parity (or reflection of the
plane of the electron gas), which is not a symmetry in the presence of
a strong magnetic field, and will cause critical behavior, or
delocalization of the electrons, for $\sigma_{xy} \in {\mathbb Z} +
1/2$.  The term is called topological because it arises from pulling
back a closed two-form (the K\"ahler form) on $G/H$. Its integral over
two-cycles therefore takes quantized values, $2\pi n i \sigma_{xy}$
with $n \in {\mathbb Z}$, and the Hall conductivity acquires the
meaning of a topological angle $\theta=2\pi\sigma_{xy}$.

To generate specific observables of the two-dimensional electron gas,
one includes sources in the Lagrangian and takes derivatives as usual.
For example, the correlator of spectral determinants $\Omega(a_0,a_1
;b_0,b_1)$ is obtained by adding a term
	$$
	L_\omega = \Lambda^2 {\rm STr} \left( \omega Q 
	- \omega \Sigma_3 \right) \;,
	$$
where $\omega = {\rm diag}(-\ln a_0,-\ln a_1,\ln b_0,\ln b_1)$ and 
$\Lambda$ is an ultraviolet cutoff.  The parameters $a_0$ and $b_0$ 
are subject to
	$$
	{\rm Re} \ln a_0 < 0 \;, \quad {\rm Re} \ln b_0 < 0 \;.
	$$
They act as regulators for the zero modes of the hyperbolic degrees of
freedom $(Z_{\rm BB})$ (the ``BB-sector'') of the matrix field $Q$.

For many years the disordered electron community has debated whether
Pruisken's model is the ``correct'' or ``good'' theory of the integer
quantum Hall plateau transition.  The issue is not whether the model
for $\sigma_{xy} = 1/2$ is massless (it certainly is), but whether on
the critical line $\sigma_{xy} = 1/2$ there exists a renormalization
group fixed point where the beta function for $\sigma_{xx}$ vanishes.
From the practical point of view, one also wants to know whether one
can analytically solve \cite{zz} such a fixed-point theory, if it
exists.  These questions have never been answered conclusively.  At
weak coupling $(\sigma_{xx} \gg 1)$, the beta function can be computed
by perturbation theory.  The topological term $L_{\rm top}$ is
perturbatively invisible, and hence the situation is the same as for
$\sigma_{xy} = 0$: the one-loop beta function vanishes, as the
geometry of the (symmetric) space $G/H$ is Ricci flat, but in two-loop
order the quantum fluctuations kick in and drive the theory towards
strong coupling.  On the basis of dilute instanton gas calculations,
Pruisken and collaborators \cite{llp} argued that nonperturbative
effects due to the topological term terminate the RG flow at some
finite coupling $\sigma_{xx}^*$.  However, close scrutiny shows that
the argument is not really convincing, as the instanton gas can only
be controlled, if at all, at weak coupling and a far extrapolation to
strong coupling is required.  Thus the status of Pruisken's model as a
candidate for the fixed-point theory has remained unclear.

However, a certain bias {\it against} Pruisken's model came from the
following observation.  When the parameters $\ln a_0$ and $\ln b_0$
are moved to minus infinity, the fields $Z_{\rm BB}$, $Z_{\rm BF}$, 
and $Z_{\rm FB}$ become massive and drop out of the theory, leaving 
behind a massless sector ${\rm S}^2 \simeq {\rm U}(2)/{\rm U}(1) 
\times {\rm U}(1)$ governed by the Lagrangian 
	$$
	L = {1 \over g^2} {\partial_\mu \bar z \, \partial_\mu z \over 
	(1 + \bar z z)^2 } + {\theta \, \epsilon_{\mu\nu} \over 2\pi} 
	{\partial_\mu \bar z \, \partial_\nu z \over (1 + \bar z z)^2 }
	+ \Lambda^2 \ln(a_1 b_1) {1 - \bar z z \over 1 + \bar
	z z} \;,
	$$
where $z \equiv Z_{\rm FF}$, $1/g^2 = \sigma_{xx}$, and $\theta =
2\pi\sigma_{xy}$.  This is the Lagrangian of the so-called ${\rm
O}(3)$ nonlinear sigma model with a topological term. It has a global
${\rm SU}(2)$ [or ${\rm O}(3)]$ symmetry, perturbed by the
symmetry-breaking field $\ln(a_1 b_1)$.  Notice that on the critical
line $\sigma_{xy} = 1/2$, the topological angle $\theta$ equals $\pi$.

There exists a general consensus about the fate under renormalization
of the ${\rm O}(3)$ model at $\theta = \pi$.  Perturbation theory
shows that the coupling constant of this field theory, like Pruisken's
model, increases under renormalization.  (It already does in one-loop
order.)  In the strong-coupling limit $g^2\to\infty$, the low-energy
Hamiltonian of the lattice-regularized theory becomes \cite{sr} the
Hamiltonian of the 1d Heisenberg model of an ${\rm SU}(2)$-invariant
quantum antiferromagnet with spin $S = 1/2$.  Recall that we arrived
at the same theory already in Section \ref{sec:cc}, by starting from
the network model and pointing out that the critical correlator
$\Omega(0,a_1;0,b_1)$ maps on the partition function of a six-vertex
model, which in turn has the Heisenberg antiferromagnet for its
(spatially) anisotropic limit.  As was mentioned earlier, the
low-energy physics of the latter is governed by the ${\rm SU}(2)_1$
WZW model, perturbed by a current-current interaction.  The
perturbation is marginally irrelevant, and the theory in the infrared
flows to a conformal invariant fixed point, where the global ${\rm SU}
(2)$ invariance of the nonlinear sigma model is promoted to an ${\rm
SU}(2)_{\rm L} \times {\rm SU}(2)_{\rm R}$ current algebra.

By analogy, one expects a similar scenario to take place for
Pruisken's model at $\sigma_{xy} = 1/2$ and $\omega = 0$: global
invariance under the complexification $G \equiv {\rm SL}(2|2)$ of
${\rm SU}(1,1|2)$ should be promoted to a chiral symmetry $G_{\rm L}
\times G_{\rm R}$ in the fixed-point theory.  The argument for
symmetry doubling will be reviewed in the next section, after the
introduction of its essential ingredient, namely a ``superspin''
analog (for Pruisken's model) of the spin-1/2 Heisenberg chain.

\section{Superspin chain}
\label{sec:spin}

We start out with a concise description of the superspin chain,
and will indicate the relation to the Chalker-Coddington network and
Pruisken's nonlinear sigma model afterwards.  The chain is a
one-dimensional ``antiferromagnet'' with degrees of freedom that take
values in an alternating sequence of ${\rm gl}(2,2)$ modules $V$ and
$V^*$.  Our first task is to describe these modules.  They were first
identified in unpublished work by N. Read \cite{read}. 

To begin, let $E_{ij}$ denote the matrix whose entries are zero
everywhere except at the intersection of the $i$-th row with the
$j$-th column where the entry is unity.  By ${\rm gl}(2,2)$ we mean
the Lie superalgebra spanned by $\{E_{ij}\}_{i,j=0,...,3}$ over
${\mathbb C}$, with the bracket or supercommutator given by
	\begin{eqnarray*}
	[E_{ij},E_{kl}] &=& E_{ij}E_{kl}-(-1)^{(i+j)(k+l)}E_{kl}E_{ij} \\
	&=& \delta_{jk} E_{il} - \delta_{li} (-1)^{(i+j)(k+i)} E_{kj} \;.
	\end{eqnarray*}
Next put
	$$
	\begin{array}{llll}
	\bar c_0 = b_{+}^\dagger \;, \quad
	&\bar c_1 = f_{+}^\dagger \;, \quad
	&\bar c_2 = - b_{-} \;, \quad
	&\bar c_3 = f_{-} \;, \\
	c_0 = b_{+} \;, \quad
	&c_1 = f_{+} \;, \quad
	&c_2 = b_{-}^\dagger \;, \quad
	&c_3 = f_{-}^\dagger \;,
	\end{array}
	$$
where $b_{\pm}^\dagger, f_{\pm}^\dagger$ and $b_\pm , f_\pm$ are
creation and annihilation operators for ``charged'' $(\pm)$ bosons and
fermions.  They obey the canonical commutation and anticommutation
relations, and act in a Fock space ${\cal F}$ with vacuum $|0\rangle$.
The mapping
	$$
	E_{ij} \mapsto S_{ij} \equiv \bar c_i c_j
	$$
determines a representation of ${\rm gl}(2,2)$ on ${\cal F}$.
According to general theory \cite{howe}, this representation nicely
decomposes as a {\it direct sum of irreducibles}.  The irreducible
representation spaces are labelled by an integer, which is the
eigenvalue of the operator $C = \sum_i \bar c_i c_i$ generating the
{\it center} of ${\rm gl}(2,2)$.  It is evident that $C$ counts the
difference between the number of positively and negatively charged
particles (or of ``retarded'' and ``advanced'' particles, using the
terminology of the disordered electron system).  Thus $C$ is the
total charge.  We want the irreducible space, $V$, on which $C$
vanishes.  Note that $V$ contains the vacuum $|0\rangle$.  The latter
actually is a lowest-weight state for the ${\rm gl}(2,2)$ module $V$.

The definition of the conjugate module $V^*$ completely parallels 
that of $V$, except that the fundamental identifications change to
	$$
	\begin{array}{llll}
	\bar c_0 = - b_{+} \;, \quad
	&\bar c_1 = f_{+} \;, \quad
	&\bar c_2 = b_{-}^\dagger \;, \quad
	&\bar c_3 = f_{-}^\dagger \;, \\
	c_0 = b_{+}^\dagger \;, \quad
	&c_1 = f_{+}^\dagger \;, \quad
	&c_2 = b_{-} \;, \quad
	&c_3 = f_{-} \;,
	\end{array}
	$$
and the vacuum state of $V^*$ is denoted by $|\bar 0\rangle$ for
better distinction.  (Equivalently, we could keep the operator
identifications, and alternate the definition of the vacuum
\cite{network}.)  Both modules $V$ and $V^*$ are infinite-dimensional,
since we can keep creating boson pairs $b_+^\dagger b_-^\dagger$ with
no limit.  They are naturally completed as Hilbert spaces with the
usual Hermitian scalar product that makes $b^\dagger$ the adjoint of
$b$, and $f^\dagger$ the adjoint of $f$.  Adopting the spin
terminology used in Section \ref{sec:cc}, we refer to the elements of
$V$ and $V^*$ as the state vectors of a ``superspin''.

For some purposes (as in Section \ref{sec:quantum}), one wants to
interpret the conjugate module $V^*$ as the linear space {\it dual} to
$V$.  This is done as follows.  We temporarily restrict the range of
$i,j$ to $i \in \{0,1\}$ and $j \in \{2,3\}$.  With this restriction,
the superspin generators $S_{ij} \equiv S_{ij}^+$ are referred to as
``raising'' operators, and $S_{ji} \equiv S_{ji}^-$ as ``lowering''
operators.  The vectors of $V$ $(V^*)$ are then created by acting with
raising (lowering) operators on $|0\rangle$ (resp. $|\bar 0\rangle$).
To identify $V^*$ as the dual space of $V$, we need to define a
nondegenerate pairing $V^* \times V \to {\mathbb C}$, $(w,v) \mapsto
\langle w , v \rangle$.  Let $v = S_{i_1 j_1}^+ ... S_{i_n j_n}^+
|0\rangle\in V$, and $w=S_{j_n^\prime i_n^\prime}^- ... S_{j_1^\prime
i_1^\prime}^- |\bar 0 \rangle \in V^*$.  We apply $w$ to $v$ by taking
all operators $S^\pm$ to act in the {\it same} module, say $V$, and
evaluating
	$$
	\langle w , v \rangle = \langle 0 | S_{j_n^\prime i_n^\prime}^-
	... S_{j_1^\prime i_1^\prime}^- S_{i_1 j_1}^+ ... S_{i_n j_n}^+
	| 0 \rangle \;.
	$$
The value of this expression is completely determined by the bracket
relations of ${\rm gl}(2,2)$, and by $S_{ji}^- | 0 \rangle = 0$ and
$S_{ii} | 0 \rangle = 0$, $S_{jj} | 0 \rangle = | 0 \rangle (-1)^{j+
1}$.  (Note, in particular, that no operation of taking an adjoint is
involved.)  For example,
	$$
	\langle 0 | S_{j^\prime i^\prime}^- S_{ij}^+ | 0 \rangle = 
	\langle 0 | [ S_{j^\prime i^\prime}^- , S_{ij}^+ ] | 0 \rangle
	= \delta_{i i^\prime} \langle 0 | S_{j^\prime j} | 0 \rangle
	= (-1)^{j+1} \delta_{i i^\prime} \delta_{j j^\prime} \;.
	$$
By linear extension, we get a bilinear form $V^* \times V \to {\mathbb
C}$, which is readily shown to be nondegenerate.  Thus, $V^*$ can be
viewed as a space of linear functions on $V$, and is therefore dual to
$V$.

We now define the superspin chain as an alternating sequence,
	$$
	... \otimes V \otimes V^* \otimes V \otimes 
	V^* \otimes V \otimes V^* \otimes ... \;,
	$$
of an even number $N$ of superspins, with Hamiltonian
	$$
	{\cal H} = \sum_n \sum_{i,j=0}^3 S_{ij}(n) (-1)^{j+1} S_{ji}(n+1)
	$$
at the critical point.  The bilinear expression $\sum_{ij} S_{ij}
(-1)^j S_{ji}$ represents the quadratic Casimir invariant of ${\rm
gl}(2,2)$.  To move the system off criticality, one staggers the
coupling between the sites $n$.  The alternation of the space of
states between $V$ and $V^*$ renders the chain ``antiferromagnetic''
in character.  Indeed, the two-superspin system $V \otimes V^*$ with
Hamiltonian ${\cal H}$ has been shown \cite{iqhe} to have a ${\rm
gl}(2,2)$ invariant ground state with zero energy (separated by a gap
from a continuum of excited states) as befits the supersymmetric
generalization of a quantum antiferromagnet.

Before carrying on, we insert the following technical remark.  The Lie
superalgebra ${\rm gl}(2,2)$ suffers from the disease of being
non-semisimple.  It has a one-dimensional center generated by the unit
matrix $I \equiv \sum_i E_{ii}$, which obviously commutes with all
matrices.  At the same time, since ${\rm STr}\;[ X , Y ] = 0$ for any
pair $X , Y \in {\rm gl}(2,2)$, there exists a one-dimensional
subspace of elements that never appear on the right-hand side of any
supercommutation relation.  This is the subspace of multiples of the
generator with nonvanishing supertrace, $F \equiv \sum_i (-1)^i
E_{ii}$.  Removal of the latter defines the subalgebra ${\rm sl}(2,2)
\subset {\rm gl}(2,2)$.  One can also take the quotient by the center
and pass to ${\rm psl}(2,2) = {\rm sl}(2,2) / {\mathbb C} \cdot I$,
which is still a Lie superalgebra.  The quadratic Casimir decomposes
as
	$$
	{\cal C}_{{\rm gl}(2,2)} = {\cal C}_{{\rm psl}(2,2)} + 
	{\rm const} \times I F \;.
	$$
Since $I$ is represented by charge $C = 0$ on $V$ and $V^*$, it is
possible to express the Hamiltonian ${\cal H}$ of the superspin chain
completely in terms of the generators of ${\rm psl}(2,2)$.  However,
for the purposes of the present section we find it convenient not to
project on ${\rm psl}(2,2)$, but rather to tolerate the presence of a
nontrivial center and work with ${\rm sl}(2,2)$.  (The center will
then give rise to unphysical gauge degrees of freedom in the field
theory.)

Given the Lie superalgebra representation $E_{ij} \mapsto \bar c_i
c_j$, we can exponentiate to obtain a ${\rm SL}(2|2)$ group action on
$V$ (or $V^*$, it works the same way in both cases) by $$ g \mapsto
\exp \left( \sum_{ij} \bar c_i \, (\ln g)_{ij} \, c_j \right) \;.  $$
For this action to make sense, we have to allow multiplication of the
state vectors of $V$ and $V^*$, now viewed as ${\rm sl}(2,2)$ modules,
by the anticommuting parameters that are needed to form the
supermatrix $g \in {\rm SL}(2|2)$.  Because the eigenvalues of $\bar
c_i c_i$ are integers, the exponential is well-defined in spite of the
multi-valuednessof the logarithm.  The complex group ${\rm SL}(2|2)$,
of course, does not act by isometries on the Hilbert spaces $V$ and
$V^*$, but there exists a bosonic subgroup ${\rm SU}(1,1) \times {\rm
SU}(2)$ which does.  In other words, $V$ and $V^*$ decompose into {\it
unitary} representations of ${\rm SU} (1,1) \times {\rm SU}(2)$.  The
pseudo-unitary subgroup ${\rm SU}(1,1|2) \subset {\rm SL}(2|2)$ (the
symmetry group of Pruisken's nonlinear sigma model) is less useful
here, as some of its odd generators do not act unitarily (or create
``states with negative norm'').  The ${\rm SL}(2|2)$ action on $V$ and
$V^*$ extends to an action of ${\rm SL}(2|2)^N$ on the superspin chain
of length $N$.  Since $\sum_{ij} S_{ij} (-1)^j S_{ji}$ is the
quadratic Casimir invariant of ${\rm gl}(2,2)$ represented on $V$ or
$V^*$, the diagonal of ${\rm SL} (2|2) ^N$ (the global action)
commutes with the superspin Hamiltonian ${\cal H}$.  This completes
our definition of the superspin chain and its symmetries.

There exist several ways of arriving at the superspin chain and its
Hamiltonian.  The first published derivation \cite{iqhe} took
Pruisken's nonlinear sigma model for its starting point.  Following
the treatment by Shankar and Read \cite{sr} of the ${\rm O}(3)$
nonlinear sigma model at $\theta = \pi$, the two-dimensional
supersymmetric field theory was discretized on a lattice of sites in
the spatial direction and put in one-dimensional quantum Hamiltonian
form.  For strong coupling, the ``large'' part of the Hamiltonian is
site-diagonal, with the single-site Hamiltonian being $(p-A)^2$, where
$p^2$ is the Laplacian on the target space ${\rm U}(1,1|2) / {\rm
U}(1|1) \times {\rm U}(1|1)$ of the nonlinear sigma model, and $A$ is
the gauge field of a fictitious magnetic monopole.  The single-site
Hamiltonian has zero-energy states, which are naturally described as
holomorphic sections of an associated line bundle.  For the case of an
even-numbered site, the vector space of these zero modes is precisely
the module $V$, and for odd sites it is $V^*$.  The degeneracy between
the zero-energy states is lifted by the coupling between sites, and
this interaction projects on the superspin Hamiltonian ${\cal H}$ written
down above.  Thus we conclude that ${\cal H}$ governs the low-energy physics
of Pruisken's model at strong coupling.

The superspin Hamiltonian can be obtained more easily by starting from
the spatially anisotropic limit of the Chalker-Coddington model, which
is a sequence of counterpropagating ``edges'' coupled by random
complex tunneling amplitudes.  Interpreting the functional integral
representation of the disorder averaged theory as a coherent-state
path integral, and passing to a Hamiltonian description with the help
of the transfer matrix, one arrives quite directly (and without having
to make approximations) at the spin chain.  This was first done using
replicas by D.-H. Lee \cite{dhlee}.  A supersymmetric version was
given in \cite{network,kondev}.

The important conclusion of all this is that Pruisken's nonlinear
sigma model at strong coupling is equivalent to the Chalker-Coddington
model, and both can be represented as a superspin chain.  Thus there
is a convergence of models and formulations, and the focus now is on
the antiferromagnetic superspin chain.  There exists some recent
numerical work using the density-matrix renormalization group, along
with analytical ideas motivated by the Lieb-Schultz-Mattis theorem,
which confirm the expectation that the superspin chain is quantum
critical \cite{marston}.  There exist also attempts \cite{gade98} to
deform the superspin chain to an integrable model that is solvable in
the Yang-Baxter sense.  Unfortunately, the technical difficulties
encountered there are enormous, and an analytical solution does not
seem to be within close reach.

Although the antiferromagnetic superspin chain at low energies is not
easy to control, we can still make some useful {\it qualitative}
predictions from it, as follows.  Recall that the action of $G = {\rm
SL}(2|2)$ on $V$ and $V^*$ gives an action of $G^N$ on the chain of
length $N$.  If $g = g_1 \times g_2\times ...\times g_N \in G^N$
(Cartesian product), we write $\hat g = \hat g_1 \times \hat g_2
\times ... \times \hat g_N$ for the element acting on the chain.  The 
ground state of the chain with Hamiltonian ${\cal H}$ is some
complicated object, $|\psi_0\rangle$.  We do not know how to describe
it analytically, but it certainly exists, and since the superspin
chain is antiferromagnetic, we can be sure that $|\psi_0\rangle$ is
{\it invariant} with respect to the global $G$ action.  (The ground
state of a $1d$ isotropic antiferromagnet with an even number of spins
is always a spin singlet.)  In formulas: $\hat g | \psi_0 \rangle = |
\psi_0 \rangle$ for any global $\hat g = \hat g_1 \times \hat g_1 
\times ... \times \hat g_1$.  The $G$ symmetry of the Hamiltonian is
expressed by the equation $\hat g {\cal H} \hat g^{-1} = {\cal H}$,
again for global $\hat g$.

Recall next that $V$ and $V^*$ are {\it irreducible} ${\rm sl}(2,2)$
modules.  Irreducibility of $V$ and $V^*$ means that {\it all} state
vectors of the superspin chain are reached by acting with $G^N = {\rm
SL}(2|2)^N$ on some reference state.  Now, what we are after is the
{\it low-energy physics} of the chain.  For a chain with a large
correlation length, the low-energy states are generated by acting with
``slowly varying'' $\hat g$ on the ground state $|\psi_0\rangle$.  In
other words, we expect the existence of a continuum limit where the
low-energy states form an irreducible module for a {\it loop group},
$LG$.  (For this we impose periodic boundary conditions on the chain).
The elements of the loop group $LG$ are smooth maps ${\rm S}^1 \to G$,
denoted by $\hat g(x)$.  The action $g \mapsto \hat g$ of $G^N$ on the
discrete chain carries over to a representation $g(x)h(x)\mapsto\hat
g(x) \hat h(x)$ of the loop group, in the continuum limit.

Given the action of the loop group, we can form $LG$-coherent states
$\hat g | \psi_0 \rangle$.  Let us now postulate the existence of a 
functional integral measure $D \hat g$, such that 
	$$
	\int D\hat g ~~ \hat g | \psi_0 \rangle 
	\langle \psi_0 | \hat g^{-1} = {\rm id} 
	$$ 
is a resolution of unity for the low-energy sector of the superspin
chain.  Using it we can pass to a $LG$-coherent state path integral
\cite{mstone}, with the functional integrand being a product of matrix
elements of the form
	\begin{equation}
	\langle \psi_0 | \hat g_{\tau + \delta \tau}^{-1} \, 
	{\rm e}^{-\epsilon {\cal H}} \hat g_\tau^{\vphantom{-1}} | \psi_0
	\rangle \;.  \label{me}
	\end{equation}
If we knew $|\psi_0\rangle$ and knew how to compute such matrix
elements, we would have a constructive way of deriving the low-energy
effective field theory of the QH plateau transition, which will be a
functional integral over continuous fields $g(x,\tau)$.  Since we know
neither, the present argument remains formal.

Nevertheless, the $LG$-coherent state path integral exists in
principle and leads us to the following assertion: the functional
integral for $g(x,\tau)$ has a {\it chiral} symmetry $G_{\rm L} 
\times G_{\rm R}$.  Indeed, since $g(x) \mapsto \hat g(x)$ is a 
representation, the matrix element (\ref{me}) does not change under
global transformations $g_\tau(x) \mapsto g_{\rm L} g_\tau(x) g_{\rm
R}$ and $g_{\tau + \delta\tau}(x) \mapsto g_{\rm L} g_{\tau + \delta
\tau}(x) g_{\rm R}$.  The left invariance is due to invariance of the 
Hamiltonian, the right invariance is a result of invariance of the
ground state.

The argument given so far can be made regardless of whether the chain
is critical or not.  (The Hamiltonian is always invariant, and so is
the ground state.)  Now comes an important distinction.  Away from
criticality, where the correlation length is finite, the system has a
{\it mass gap}.  Therefore, the low-energy phase space must be reduced
in some manner.  Technically speaking, the loop group does not act
freely ({\it i.e.}, not all states created by its generators are
linearly independent), and hence some its degrees of freedom need to
be gauged out in the functional integral.  Gauging happens on the
right (where the loop group acts on states) and thus interferes with
the symmetry action on the right, leaving only the left action as a
good symmetry of the theory.  (This is the situation in Pruisken's
nonlinear sigma model.)  On the other, in the massless theory we do
expect the loop group to act freely, modulo the global invariance
$\hat g | \psi_0 \rangle = | \psi_0 \rangle$ leading to the phase
space being $LG / G$.  In that case the naive $G_{\rm L} \times G_{\rm
R}$ invariance of the coherent-state path integral should exist as a
true symmetry of the properly defined field theory.  This then is our
main conclusion: in the fixed-point theory, the global $G$ symmetry of
Pruisken's model is promoted to a {\it larger} symmetry, namely an
invariance under two copies of the symmetry group, $G_{\rm L} \times
G_{\rm R}$.

To avoid confusion, let me emphasize that we are not arguing in favor
of a stronger statement due to Affleck \cite{affleckprl}.  That
argument says that for {\it unitary} theories with continuous $G$
symmetry the fixed point acquires an affine Lie algebra symmetry,
which is {\it local} and hence infinite-dimensional.  All we get from
the above argument is a chiral doubling of the {\it global} symmetry
group.  (In order for the symmetry to become affine, the field
$g(x,\tau)$ would have to separate into left-moving and right-moving
waves.  We will see later that this does not happen in the present
case.)

There exists a second piece of valuable information we can infer from
the $LG$-coherent state path integral of the superspin chain.  Recall
that $g \in {\rm SL}(2|2)$ acts on $V$ and $V^*$ by $\exp \sum_ {ij}
\bar c_i \, (\ln g)_{ij} \, c_j$.  Evidently the multiples of unity,
$g = {\rm e}^s \cdot 1_4$, couple to the total charge $C=\sum_i\bar
c_i c_i$.  By construction, $C = 0$ on every state vector of $V$ and
$V^*$, and therefore $\sum_i \bar c_i(n) c_i(n)$ {\it vanishes on all
states} of the superspin chain.  As a result, nothing depends on the
coherent state parameter $s_\tau(n)$ conjugate to the local charge
$\sum_i\bar c_i(n)c_i(n)$.  In the continuum limit, this independence
becomes an invariance of the $LG$-coherent state path integral under
local gauge transformations $g(x,\tau) \mapsto {\rm e}^{s(x,\tau)}
g(x,\tau)$ where ${\rm e}^{s(x,\tau)} \in {\rm GL}(1)$ is any
invertible multiple of the unit matrix.  This symmetry can be seen to
be a direct consequence of the local ${\rm U}(1)$ gauge invariance of
the Chalker-Coddington model [${\rm GL}(1)$ here is precisely the
complexification of ${\rm U}(1)$].

\section{Constraints on the fixed-point theory}
\label{sec:constraints}

We have reviewed the argument why we are going to abandon Pruisken's
model and look for another field theory with manifest conformal
invariance.  Ideally, we would like to deduce the theory {\it
constructively}, by starting from a model of disordered electrons such
as the network model of Chalker and Coddington, and making controlled
approximations.  Unfortunately, this looks like a rather difficult if
not impossible project, as the fixed point always appears to be at
strong coupling, no matter what choice of starting point is made (the
network model, the nonlinear sigma model, or a model of disordered
Dirac fermions \cite{lfsg}).  In such a situation, we are forced to
resort to indirect reasoning.  The viable procedure is to make an
educated guess, and verify its correctness by comparing the
consequences to known results.  In making such a guess, we are guided
by the fact that the theory we are looking for meets the following
list of requirements.

1. The field theory is defined by a stable functional integral with a
target space of Euclidian signature (as opposed to Lorentzian or other
signature).  This condition eliminates, in particular, target spaces
such as ${\rm PSL}_{\mathbb R}(2|2)$ or ${\rm PSU}(1,1|2)$, which have
appeared in recent work motivated by string propagation in ${\rm
AdS}_3$ backgrounds \cite{bvw,bzv}.

2. Conformal invariance of the fixed-point theory is manifest.  The
energy-momentum tensor splits into a holomorphic and an
antiholomorphic piece, the Fourier components of which obey the
commutation relations of a Virasoro algebra.  (Note that it is {\it
not} obligatory for the theory have an affine Lie algebra symmetry.)

3. The partition function of the theory is normalized to unity: $Z =
1$, independent of the size or other parameters of the system.  For a
conformal field theory this implies that the Virasoro central charge
$c$ vanishes.  An immediate consequence is that the theory cannot be
unitary, as this would require $c > 0$.

4. The field-theory representation of the correlation function
(\ref{speccorr}) reduces to unity on setting either $a_0 = a_1$ or
$b_0 = b_1$.  More generally, correlation functions and other
observables of the theory become trivial when only one causal sector
(retarded or advanced) is probed.

5.  The Hamiltonian of the antiferromagnetic superspin chain is
invariant under a global action of $G = {\rm SL}(2|2)$.  In the
fixed-point Lagrangian, this invariance is promoted to a (global)
chiral symmetry $G_{\rm L} \times G_{\rm R}$.  Invariance under the
center of $G$, which is the subgroup generated by the unit matrix, is
present as a {\it local} gauge symmetry.

6. Massive perturbation of the Boson-Boson sector (or equivalently,
sending the parameters $\ln a_0$ and $\ln b_0$ to minus infinity)
preserves criticality of the theory, by leaving some of the bosonic
fields massless.  The reduced conformal field theory for the massless
modes is the ${\rm SU}(2)_1$ WZW model.

7. The theory contains an operator corresponding to the density of
states of the disordered electron system, the scaling dimension of
which equals zero.  This requirement follows from the known fact
\cite{wegnerDOS} that the density of states at the critical point 
under consideration is noncritical.

8. The fixed-point theory reproduces the critical exponents for
various correlation functions that are known from numerical and real
experiments.

In the sequel, we will describe a field theory which demonstrably
satisfies all of the requirements 1--7.  Since the theory turns out to
be of a novel kind, with solutions not being readily available,
requirement 8 can only be partially verified at the present time and
will need further work in the future.

\section{Symmetric superspaces and cs-manifolds}
\label{sec:motiv}

The correct definition of the target space of the field theory relies
on the two notions of Riemannian symmetric superspace and cs-manifold.
They are not entirely standard and are not commonly understood.  It is
therefore proper to put the general development on hold, and motivate
and explain these notions of supermanifold theory in some detail,
which is what we will do below.

The base of the supersymmetric target space $G / H$ of Pruisken's
model is ${\rm H}^2 \times {\rm S}^2$.  As we recalled, the nonlinear
sigma model with target space ${\rm S}^2$ (the ${\rm O}(3)$ model) and
topological angle $\theta = \pi$ flows under renormalization to a
theory of fields taking values in ${\rm SU}(2) \simeq {\rm S}^3$.
Thus quantum fluctuations in this case promote the two-sphere to the
three-sphere.  From that we take it that we should try promoting the
noncompact sector ${\rm H}^2$ of Pruisken's model to a
three-hyperboloid ${\rm H}^3$.

The first question to address then is whether there exist conformal
field theories of ${\rm H}^3$-valued fields.  The answer is yes.
${\rm H}^3$ is diffeomorphic to the noncompact symmetric space ${\rm
SL}(2,{\mathbb C})/{\rm SU}(2)$, which in turn can be identified with
the positive Hermitian unimodular $2 \times 2$ matrices $M$, by
setting $M = g g^\dagger$ and letting $g$ run through ${\rm SL}(2,
{\mathbb C})$.  For the latter we can write down a WZW Lagrangian,
	$$
	L = -(|k|/2\pi)\,{\rm Tr}\,\partial M\,\bar\partial M^{-1}+(ik 
	/12\pi)\,{\rm d}^{-1}{\rm Tr}\,(M^{-1}{\rm d}M)^{\wedge 3} \;.
	$$
The partial derivatives $\partial$ and $\bar\partial$ are the usual
ones, {\it i.e.}, if $z$ and $\bar z$ are complex coordinates for
two-dimensional space (or space-time), we require $\partial z =
\bar\partial \bar z = 1$ and $\partial\bar z = \bar\partial z = 0$.
The notation ${\rm d} ^{-1} \Omega$ means any potential of the 3-form
$\Omega$.  In contrast with the compact version, where $k$ is
topologically quantized, the coupling $k$ here need not be an integer,
as ${\rm H}^3$ is diffeomorphic to ${\mathbb R}^3$, which has trivial
cohomology.  Another difference is that the transformation group ${\rm
SU}(2)_{\rm L} \times {\rm SU}(2)_{\rm R}$ acts on ${\rm SU}(2)$
independently on the left and right, whereas $g \in {\rm SL}(2,
{\mathbb C})$ acts on $M$ by $M \mapsto g M g^\dagger$, with the right
factor being the Hermitian conjugate of the left one.  Thus the left
and right actions are tied to each other.  This distinction disappears
at the level of infinitesimal symmetries, since we can always pass
with impunity to the complexified tangent space, thereby making the
left and right actions independent.
	
Noncompact WZW models of the above type play a role in the functional
integral version of the coset construction \cite{gk}.  They and
related models have been studied by high-energy physicists
\cite{dvv,ks} to a certain extent, but are notoriously hard to solve
completely, because of the complications that come from having to deal
with the representation theory of a noncompact group.  Nevertheless,
they do exist as theories with a stable functional integral, and this
is all we are require for now.  It is reasonable to expect the above
Lagrangian to be a building block of the supersymmetric theory we wish
construct.

The next question is: can we combine ${\rm H}^3$ and ${\rm S}^3$ into
a supersymmetric target space, giving rise to a WZW-type field theory
with the properties listed in Section \ref{sec:constraints}?  Again,
the answer will turn out be yes.  The present paper was triggered by
the appearance \cite{bvw,bzv} of two superstring-related articles
revolving around ${\rm AdS}_3 \times {\rm S}^3$, where ${\rm AdS}_3$
denotes three-dimensional anti-deSitter space.  For our purposes,
${\rm AdS}_3$ is identified with the manifold of the group ${\rm
SU}(1,1)$, which preserves a quadratic form of Lorentzian signature
$(+ + - )$, and has one compact and two noncompact directions. Thus
the natural geometry of ${\rm SU}(1,1)$ is non-Riemann.  It is not
hard to see that the product of ${\rm SU}(1,1)$ with ${\rm SU}(2)
\simeq {\rm S}^3$ forms the bosonic subgroup of a Lie supergroup ${\rm
PSU}(1,1|2)$, obtained from ${\rm U}(1,1|2)$ by requiring unit
superdeterminant and factoring out the ${\rm U}(1)$ generated by the
unit matrix.  The important message from \cite{bvw,bzv} is that the
nonlinear sigma model with target space ${\rm PSU}(1,1|2)$ is
conformal for {\it any value of the coupling}.  (In fact, this message
already follows from the work of Gade and Wegner \cite{gw}.)  The last
feature looks very attractive from our perspective, as it suggests
enough flexibility to accommodate the strange critical exponents that
surfaced in recent numerical work \cite{jmz}.

Needless to say, the ${\rm PSU}(1,1|2)$ nonlinear sigma model is {\it
not} the field theory we want.  Requirement 1 of our list says that
target spaces with a non-Riemannian metric are not acceptable.  Coming
from statistical physics, we insist on having a field theory with an
action functional bounded from below.  The natural remedy, of course,
is to trade ${\rm PSU}(1,1|2)$ for a related space with the required
Euclidian, or definite, signature.  Such a variant, with base ${\rm
H}^3 \times {\rm S}^3$, was mentioned in \cite{bvw}, but was quickly
dismissed with the statement that ``there is apparently no change of
variables that makes the resulting couplings to fermions real''.  I
fully agree with the statement that there are problems with reality
concerning the fermions, and these are inevitable.  Indeed, the
couplings would be real if the target space arose as the stable set of
some antilinear involutory automorphism of the complexified group
${\rm PSL}(2|2)$.  For the case of ${\rm PSU}(1,1|2)$, such an
automorphism is easily constructed, by adaptation from the definition
of ${\rm U}(1,1|2)$ (see Section \ref{sec:nlsm}).  However, if the
stable set is to be based on the manifold ${\rm H}^3 \times {\rm
S}^3$, one can show that no such automorphism exists.\footnote{Let me
digress and mention in passing that the field of disordered electron
systems began to battle with non-reality of fermions in 1992, when
Gade \cite{gade} discovered new universality classes in sublattice
models with symmetries that are now called ``chiral''.  It was
initially thought that no supersymmetric field-theory representation
of such systems exists.  The solution to the puzzle was contained in a
paper by Andreev, Simons, and Taniguchi \cite{ast} who noted that
reality of the fermions, or of the couplings to fermions, was
dispensible from a statistical physics point of view.  This idea was
pursued in \cite{suprev} where complex conjugation of fermions was
abandoned altogether, and the notion of Riemannian symmetric
superspace was formulated.}

The point to be stressed here is that the superspace we are after,
namely a variant of ${\rm PSU}(1,1|2)$ with Euclidian signature, is
not a supermanifold in the usual sense: it exists neither as a
real-analytic nor as a complex-analytic supermanifold. (A
supermanifold is called ${\mathbb K}$-analytic if the transition
functions between its coordinate charts are ${\mathbb K}$-analytic.)
Rather, it belongs to a category of objects called {\it Riemannian
symmetric superspaces} in \cite{suprev}.  The distinct nature of these
and more general objects has also been noted in recent lectures on
supersymmetry by J. Bernstein \cite{bernstein}.  He has drawn
attention to a large category of superspaces which he calls {\it
cs-manifolds} (``c'' for complex, and ``s`` for super).  Since the
notions of Riemannian symmetric superspace and/or ${\rm cs}$-manifold
may be unfamiliar, we will now elaborate somewhat, not digging into
the foundations of the subject, but reviewing the main idea and
illustrating it with a well-chosen example.

According to \cite{suprev}, a Riemannian symmetric superspace is a
highly structured object consisting of the following data.  First of
all, we are given a Lie supergroup $G_{\mathbb C}$ and a subgroup
$H_{\mathbb C}$.  Both $G_{\mathbb C}$ and $H_{\mathbb C}$ are {\it
complex}, so their quotient $G_{\mathbb C} / H_{\mathbb C}$ is
naturally a complex-analytic supermanifold.  Secondly, by fixing a
nondegenerate invariant quadratic form on ${\rm Lie}(G_{\mathbb C})$
(we assume such a form exists), the homogeneous space $G_{\mathbb C} /
H_{\mathbb C}$ is equipped with a $G_{\mathbb C}$-invariant
supersymmetric second-rank tensor, $\kappa$.  We say that $G_{\mathbb
C} / H_{\mathbb C}$ carries a ``supergeometry''.  Thirdly, the
supermanifold $G_{\mathbb C} / H_{\mathbb C}$ is based on a complex
manifold $\left(G_{\mathbb C} / H_{\mathbb C}\right)_0$.  From it we
select a {\it real} submanifold, $M$, with half the dimension of
$\left( G_{\mathbb C} / H_{\mathbb C} \right)_0$.  $M$ inherits a
geometry from the supergeometry of $G_{\mathbb C} / H_{\mathbb C}$ by
restriction.  With these provisions, we call the triple $\left(
G_{\mathbb C} / H_{\mathbb C}, \kappa , M \right)$ a Riemannian
symmetric superspace (or symmetric superspace for short) if $M$ is a
Riemannian symmetric space in the classical sense \cite{helgason}.

The merit of the above construction is that it readily produces
``nice'' superspaces, while avoiding any operation of complex
conjugation or adjoint for the fermions.  We mention in passing that
there exist ten large families of symmetric superspaces, and all of
these arise in the study \cite{suprev} of random matrix statistics
using Efetov's method.

The simplest nontrivial example derives from ${\rm GL}(1|1)$, the Lie
supergroup of invertible $2 \times 2$ supermatrices $g$,
	\begin{equation}
	g = \pmatrix{a &\alpha\cr \beta &b\cr} \;.	
	\label{matrixg}
	\end{equation}
We may regard ${\rm GL}(1|1)$ as a homogeneous space $G_{\mathbb C} /
H_{\mathbb C}$ constructed by quotienting $G_{\mathbb C} = {\rm GL}
(1|1)_{\rm L} \times {\rm GL}(1|1)_{\rm R}$ by its diagonal
$H_{\mathbb C} = {\rm GL}(1|1)$.  The natural geometry on ${\rm
GL}(1|1)$ is given by $\kappa = {\rm STr} \, (g^{-1}{\rm d}g)^2$.  For
the real submanifold $M$ of the base ${\rm GL}(1, {\mathbb C}) \times
{\rm GL}(1,{\mathbb C})$ we take ${\mathbb R}^+\times {\rm S}^1$, the 
abelian group of matrices
	$$
	g = \pmatrix{{\rm e}^x &0\cr 0 &{\rm e}^{iy}\cr} \;,
	$$
where $x \in {\mathbb R}$, and $y \in [0,2\pi]$.  We will use the
suggestive notation ${\rm H}^1$ for ${\mathbb R}^+$.  The restriction
of $\kappa$ to ${\rm H}^1 \times {\rm S}^1$ is ${\rm d}x^2 + {\rm
d}y^2$, which has the attractive feature of being Riemann (or, rather,
Euclidian).  Thus ${\rm H}^1 \times {\rm S}^1$ is a Euclidian space in
the geometry inherited from supergeometry, and the triple $\left( {\rm
GL}(1|1) , {\rm STr} \, (g^{-1}{\rm d}g)^2 , {\rm H}^1 \times {\rm
S}^1 \right)$ belongs to our general category of Riemannian symmetric
superspaces.  This example indicates a general feature, namely that
the real manifold $M$ of a symmetric superspace is a product of
spaces, $M = M_{\rm B} \times M_{\rm F}$, with $M_{\rm F}$ being
compact and $M_{\rm B}$ noncompact.  (One of the factors may of course
be trivial.)

An important fact about Riemannian symmetric superspaces is that they
admit an invariant Berezin integral (or superintegral) with nice
properties.  A brief account is as follows. The ``good'' integrands
are (super-)functions on $G_{\mathbb C} / H_{\mathbb C}$ that are
holomorphic in a neighborhood of $M$.  Let $f$ denote such a function.
By the principles of supergeometry, the complex-analytic supermanifold
$G_{\mathbb C} / H_{\mathbb C}$ comes with a holomorphic Berezin form,
$\omega$, which is a rule for converting $f$ into a holomorphic
top-form $\omega[f]$ on $\left( G_{\mathbb C} / H_{\mathbb C}
\right)_0$.  The latter can be integrated over $M$ in the usual sense,
to produce the number $\int_M \omega[f]$.  Thus, the Berezin integral
is defined to be the two-step process
	$$
	f \mapsto \omega[f] \mapsto \int_M \omega[f] \;.
	$$
The first step, called the Fermi integral or ``integration over the
Grassmann variables'', is benign and needs no notion of complex
conjugation or reality of the fermions.  Complex conjugation enters
only in the second step (after the fermions have been integrated out),
in order to fix the Riemannian manifold $M$.  Thus, reality of the
fermions, or even reality of the couplings to fermions, is not an
issue here.  Put in stronger terms, reality of the fermions is an
ill-conceived and redundant concept which might as well be abolished,
from our statistical physics perspective.\footnote{For this reason, I
have banned the use of real-form supergroups such as ${\rm U}(1,1|2)$
from all of my recent work; cf. the remark made in the second
paragraph of Section \ref{sec:nlsm}.}

We turn again to our simple ${\rm GL}(1|1)$ example for illustration.
If the elements of ${\rm GL}(1|1)$ are written as in (\ref{matrixg}),
the invariant holomorphic Berezin form is easily computed to be
$\omega \equiv Dg = i^{-1} {\rm d}a \wedge {\rm d}b \, \partial^2 / 
\partial\alpha\partial\beta$, from the invariant supergeometry given 
by ${\rm STr}\,(g^{-1}{\rm d}g)^2$.  The choice of normalization
constant $i^{-1}$ is a matter of convention.  A quick check on the
expression for $Dg$ is provided by invariance under scale
transformations: since ${\rm d}(sa) = s\,{\rm d}a$ and $\partial /
\partial(s\alpha) = s^{-1} \partial/\partial\alpha$, the factor $s$
drops out as required.  One might now think that the volume integral
$\int Dg$ vanishes, as the fermionic derivatives $\partial^2 /
\partial\alpha\partial\beta$ have nothing to act on.  If this were
really so, we would have a normalization problem to solve.  In the
field-theoretic setting of a nonlinear sigma model with target space
${\bf X} \equiv \left( {\rm GL}(1|1) , \kappa , {\rm H}^1 \times {\rm
S}^1\right)$, there would be a problem with the zero modes of the
theory.  However, the conclusion $\int Dg\,{\buildrel ? \over =}\,0$
is premature and, in fact, untenable.  Because the first factor in
${\rm H}^1 \times {\rm S}^1$ is noncompact, we are well advised to
insert a regularization factor and define the volume by
	$$
	{\rm vol}({\bf X}) = \lim_{\epsilon\to 0} \int\limits_{{\rm H}^1
	\times {\rm S}^1} Dg \, \exp \left( - \epsilon \, {\rm STr}
	\, (g + g^{-1}) \right) \;.
	$$
Note that, since there are two roads to infinity on ${\rm H}^1$ ($a
\to +\infty$ and $a \to 0$), a factor ${\rm e}^{-\epsilon \, {\rm STr}
\, g}$ will not do for regularization and we really need ${\rm
e}^{-\epsilon\,{\rm STr}(g+g^{-1})}$.  An easy calculation using $Dg =
i^{-1} {\rm d}a \wedge {\rm d}b \, \partial^2 / \partial \alpha
\partial \beta$ shows that the limit $\epsilon\to 0$ exists, and 
	$$ 
	{\rm vol}({\bf X}) = 2\pi \;.  
	$$ 
On the other hand, we could have considered the compact Lie supergroup
${\rm U}(1|1)$ or, more generally, ${\rm U}(m|n)$.  In Berezin's book
\cite{berezin} one can find a demonstration that the Berezin-Haar
integral over all these groups vanishes identically:
	$$ 
	{\rm vol} \left( {\rm U}(m|n) \right) 
	= \int\limits_{{\rm U}(m|n)} Dg = 0 \;.  
	$$
This is true in particular for ${\rm U}(1|1)$, in which case the
statement can be verified by elementary means.  Furthermore, since
${\rm U}(m|n)$ is compact, regularization has a negligible effect
here.  (Incidentally, by employing the advanced machinery of
localization on supermanifolds \cite{sz}, one can relate the vanishing
of the Berezin-Haar integral to the indefinite metric of ${\rm
U}(m|n)$.)

The point we are trying to score here is this: the nonzero
normalization integral (${\rm vol} \not= 0$), along with the
Riemannian geometry of its base, distinguishes the symmetric
superspace $\left( {\rm GL}(1|1) , {\rm STr}\,(g^{-1}{\rm d}g)^2 ,
{\rm H}^1\times {\rm S}^1 \right)$ in a subtle way from its mundane
counterpart ${\rm U} (1|1)$.  The same subtle but significant
differences exist between the Lorentzian space ${\rm PSU}(1,1|2)$ and
its Riemannian variant to be constructed below.

For mathematical perspective we conclude the section with the
following elaboration.  It has to be admitted that there is much
redundancy in our setup for symmetric superspaces.  Our declared
purpose is to integrate, and the integral we want is over the real
manifold $M$ plus the fermionic ``fuzz'' surrounding it.  A concise
mathematical formulation would therefore aim to eliminate the
``parent'' space $G_{\mathbb C} / H_{\mathbb C}$ and put the focus on
$M$ and a (sheaf of) algebra(s) of superfunctions on $M$.  This can be
done.  Since no operation of complex conjugation of the fermions is
available (and yet the ``bosons'', {\it i.e.}, the coordinates of $M$,
are real), the object that arises does not exist as a real-analytic or
complex-analytic supermanifold.  Rather, what one obtains is a {\it
cs-manifold} in the terminology of Bernstein \cite{bernstein}.  The
local model for such a manifold is a graded commutative algebra of
functions on $U \subset {\mathbb R}^p$ with values in a complex
Grassmann algebra $\Lambda {\mathbb C}^q$.  As usual, the manifold is
assembled by gluing together the local charts by means of transition
functions.  

To illustrate, we return to our simple example.  We will give another
description of the symmetric superspace ${\bf X} = ({\rm GL}(1|1) ,
{\rm STr} (g^{-1}{\rm d}g)^2, {\rm H}^1 \times {\rm S}^1)$, now
reduced to a cs-manifold.  For that purpose, let us introduce a local
coordinate system (or ``superdomain'') by the Cayley map:
	$$
	g = {1 + X \over 1 - X} \;, \qquad
	X = \pmatrix{x &\xi\cr \eta &iy\cr} \;,
	$$
where $x > 1$ parametrizes ${\mathbb R}^+ \simeq {\rm H}^1$, and $y
\in {\mathbb R}$ is a local coordinate for the circle ${\rm S}^1$.
Since ${\rm S}^1$ is diffeomorphic not to ${\mathbb R}$, but to
${\mathbb R}$ with a point added at infinity, we need a second
superdomain for an atlas of the cs-manifold.  This is constructed by 
setting 
	$$
	g = \sigma_3 \, {1 + X^\prime \over 1 - X^\prime} \;, \qquad
	X^\prime = \pmatrix{x^\prime &\xi^\prime\cr \eta^\prime 
	&iy^\prime \cr} \;,
	$$
with $\sigma_3 = {\rm diag}(1,-1)$.  The transition functions 
connecting the two superdomains follow from the equations
	$$
	{1 + X \over 1 - X} = g = \sigma_3 \, {1 + X^\prime \over
	1 - X^\prime} \;.
	$$
After a little algebra, one finds
	$$
	\begin{array}{ll}
	x^\prime = x - \xi\eta/iy \;, \quad &\xi^\prime = \xi/iy \;, \\
	\eta^\prime = - \eta/iy \;, \quad &y^\prime = - 1/y \;.
	\end{array}
	$$
Simply put, the cs-manifold ${\bf X}$ can now be described as an
algebra of ``nice'' functions of $x,y,\xi,\eta$ (and, on switching
superdomains, of $x^\prime, y^\prime, \xi^\prime, \eta^\prime$).  Note
the peculiar feature that the transition functions are complex-valued,
in spite of the fact that the even coordinates $x,y$ and $x^\prime,
y^\prime$ are designed as real variables.  Moreover, there exists no
globally consistent way of imposing relations (such as $\xi=\bar\eta
/i$, say) to make them real (try it!).  This is characteristic of a
cs-manifold: although its bosons are real, we have no choice but to
leave the fermions complex and work with an algebra of complex-valued
functions.

The advantage of reducing the symmetric superspace to a cs-manifold is
economy of the mathematical structures used.  On the negative side,
the dearth of intrinsic structure to a (bare) cs-manifold make
specifying a supergeometry, a necessary prerequisite for our purposes,
a more involved procedure.  Also, harmonic analysis and some
symmetry-related aspects of the field-theoretic setting, are more
transparent in the Riemannian symmetric superspace than in the
cs-manifold picture.  We will therefore hold on to the symmetric
superspace setup.  The best approach is to use {\it both}
descriptions, the symmetric superspace and the cs-manifold, and pass
between them freely to adapt to the changing computational needs.
This is the approach we adopt in the sequel.

\section{The target manifold}
\label{sec:target}

After these preparations, we are ready to get to the point and define
the target space of the conformal field theory to be constructed.  We
will describe the target space first as a symmetric superspace and
then as a ${\rm cs}$-manifold.

Our starting point is ${\rm GL}(2|2)$, the complex Lie supergroup of
invertible $4\times 4$ supermatrices, denoted by $g$.  This space is
too large for our purposes.  (The nonlinear sigma model for it
generically has a critical density of states, in violation of
requirement 7 of our list.)  Imposing the condition of unit
superdeterminant, ${\rm SDet}(g) = 1$, reduces the complex dimension
by one.  The resulting supergroup, ${\rm SL}(2|2)$, is not semisimple,
since its Lie algebra still contains an abelian ideal consisting of
the multiples of the unit matrix.  To gain semisimplicity, we pass to
the factor group ${\rm PSL}(2|2)$, obtained by identifying in ${\rm
SL}(2|2)$ the matrices that differ only by a scalar factor.  The
elements of ${\rm PSL}(2|2)$, which are no longer matrices but are
{\it equivalence classes} of matrices, are still denoted by $g$.  We
may view ${\rm PSL}(2|2)$ as a homogeneous space $G_{\mathbb C} /
H_{\mathbb C}$, by setting $G_{\mathbb C} = {\rm PSL}(2|2)_{\rm
L}\times {\rm PSL}(2|2)_{\rm R}$.  The projection $G_{\mathbb C} \to
G_{\mathbb C} / H_{\mathbb C}$ is done via the map $(g_{\rm L} ,
g_{\rm R}) \mapsto g_{\rm L}^{\vphantom{-1}} g_{\rm R}^{ -1}$, which
divides out the diagonal $H_{\mathbb C} = {\rm PSL}(2|2)$ $(g_{\rm L}
= g_{\rm R})$.

The Lie algebra (not the Lie superalgebra) of ${\rm PSL}(2|2)$ is a
complex superspace, ${\rm psl}(2|2)$, of dimension $(6,8)$, which is
to say there are 6 complex bosons and 8 complex fermions.\footnote{The
general theory \cite{berezin} instructs us to distinguish between a
Lie superalgebra, and an associated Lie algebra with Grassmann
structure.  The former is a ${\mathbb Z}_2$-graded linear space ${\cal
G} = {\cal G}_0 + {\cal G}_1$ with a superbracket or supercommutator
defined on it.  ${\cal G}$ will usually be realized by real or complex
matrices, and there are no Grassmann variables involved.  The latter,
the Lie algebra with Grassmann structure, is obtained by picking a
large enough Grassmann algebra $\Lambda = \Lambda_0 + \Lambda_1$ and
taking the even part of the tensor product: ${\cal G}(\Lambda) =
\Lambda_0 \otimes {\cal G}_0 + \Lambda_1 \otimes {\cal G}_1$.  If
${\cal G}$ is realized by matrices, the elements of ${\cal G}(
\Lambda)$ are supermatrices, with commuting entries on the even
blocks, and anticommuting entries on the odd blocks.  Depending on the
context, the generators of $\Lambda$ are either viewed as {\it
parameters}, or take the role of odd {\it coordinates}.  For present
purposes, we denote the Lie superalgebra by ${\rm psl}(2,2)$, and the
Lie algebra with Grassmann structure by ${\rm psl}(2|2)$.}
Supergeometry is introduced by fixing on ${\rm psl}(2|2)$ a
supersymmetric quadratic form which is both nondegenerate and
invariant under the adjoint action of the group.  The Killing form
$\langle X , Y \rangle = {\rm STr} \, {\rm ad}(X) {\rm ad}(Y)$ will
not do for that purpose, as it is degenerate in the present case
\cite{kac}.  However, we can start from the Lie algebra of ${\rm SL}
(2|2)$ in the fundamental matrix representation and take
	$$
	\langle X , Y \rangle = {\rm STr} \, XY \;.
	$$
This descends to a well-defined quadratic form on ${\rm psl}(2|2)$,
since 
	$$
	\langle X + s \cdot 1_4 , Y + t \cdot 1_4 \rangle 
	= \langle X , Y \rangle
	$$ 
is independent of the scalars $s$ and $t$.  We denote the resulting
quadratic form still by $\langle X , Y \rangle$.  This form enjoys the
properties of nondegeneracy and invariance under the adjoint action of
${\rm PSL}(2|2)$.

By virtue of ${\rm PSL}(2|2)$ being a homogeneous space, the quadratic
form $\langle \bullet , \bullet \rangle$ induces an invariant
second-rank tensor (or ``metric'') $\kappa$ on ${\rm PSL}(2|2)$ by
left translation.  Given two tangent vectors at $g$, we simply move
them to the Lie algebra ${\rm psl}(2|2)$ by parallel translation with
$g^{-1}$, and then evaluate their inner product using $\langle \bullet
, \bullet \rangle$.  We denote the resulting tensor by
	$$
	\kappa = \langle g^{-1}{\rm d}g , g^{-1}{\rm d}g \rangle \;.
	$$

By setting the fermionic degrees of freedom of ${\rm PSL}(2|2)$
to zero, we obtain the six-dimensional complex manifold ${\rm PSL}
(2|2)_0$.  If ${\mathbb C}^\times$ denotes the group of invertible
complex numbers, ${\rm PSL}(2|2)_0$ consists of equivalence classes,
	$$
	{\mathbb C}^\times \, {\rm diag}(A,B) \subset {\rm PSL}(2|2) \;,
	$$
parametrized by complex $2 \times 2$ matrices $A$ and $B$, which are 
chosen to satisfy 
	$$
	{\rm Det}(A) = {\rm Det}(B) = 1 \;,
	$$
and hence lie in ${\rm SL}(2,{\mathbb C})$.  The construction of the
Riemannian symmetric superspace is completed by specifying a {\it
real} submanifold $M$ of ${\rm PSL}(2|2)_0$.  In keeping with the
discussion of Section \ref{sec:motiv}, we choose $M \simeq M_{\rm B}
\times M_{\rm F} \simeq {\rm H}^3 \times {\rm S}^3$.  In precise
terms, we take the elements of $M$ to be of the form ${\mathbb C}
^\times {\rm diag}(A,B)$ where $A$ runs through the positive Hermitian
$2 \times 2$ matrices of unit determinant: $A = h h^\dagger$ where $h
\in {\rm SL}(2,{\mathbb C})$, and $B$ runs through ${\rm SU}(2)$.  The
former set $(M_{\rm B})$ is isomorphic to ${\rm SL}(2, {\mathbb C}) /
{\rm SU}(2) \simeq {\rm H}^3$, and the latter to ${\rm S}^3$.

The elements of the tangent space of $M$ at the group identity are
represented by pairs ${\rm diag}(a,b)$ with $a$ Hermitian and $b$ 
skew-Hermitian.  Hence we have
	$$
	\left\langle {\rm diag}(a,b) , {\rm diag}(a,b) 
	\right\rangle = {\rm Tr} \, a^2 - {\rm Tr} \, b^2 = 
	{\rm Tr} \, a^\dagger a + {\rm Tr} \, b^\dagger b \ge 0 \;,
	$$
and $M \simeq {\rm H}^3 \times {\rm S}^3$ is {\rm Riemann} in the
geometry inherited from the supergeometry of ${\rm PSL}(2|2)$.  Thus
$M$ is a Riemannian symmetric space, and the triple $({\rm PSL}(2|2),
\kappa , M)$ is a Riemannian symmetric superspace.  Adopting standard
terminology \cite{kac}, we call it type $A_1|A_1$;  in symbols: 
${\bf X}_{A_1|A_1}$. 

As was discussed in the closing paragraph of Section \ref{sec:motiv},
the definition of a Riemannian symmetric superspace carries some
redundancy.  If our purpose is to define a nonlinear sigma model, we
only need the Riemannian manifold $M \simeq {\rm H}^3 \times {\rm
S}^3$ and the fermionic fuzz surrounding it, together with the
invariant supergeometry induced from ${\rm PSL}(2|2)$.  Hence, a more
concise description of the object at hand exploits the notion of ${\rm
cs}$-manifold, which is our next topic.  We can describe the ${\rm
cs}$-manifold from the coordinate point of view, by constructing an
atlas of superdomains and their transition functions.  Since ${\rm
H}^3$ is diffeomorphic to ${\mathbb R}^3$, and the second factor of
$M$ is a (three-)sphere, the minimal atlas consists of two
superdomains.  We define the first of these by using the exponential
map centered around the group unit of ${\rm PSL}(2|2)$.  In detail, we
proceed as follows.  We set
	$$
	X = \pmatrix{a &\alpha\cr \beta &b\cr} = 
	\pmatrix{a_3 &a_1-ia_2 &\alpha_{11} &\alpha_{12}\cr
		 a_1+ia_2 &-a_3 &\alpha_{21} &\alpha_{22}\cr
		 \beta_{11} &\beta_{12} &ib_3 &ib_1+b_2\cr
		 \beta_{21} &\beta_{22} &-ib_1-b_2 &-ib_3\cr} \;,
	$$
and take the even coordinates $a_1,a_2,a_3,b_1,b_2,b_3$ to be real.  
Note $a \in i{\rm su}(2)$, and $b \in {\rm su}(2)$.  Note also that
{\it no conditions of any kind} are imposed on the odd coordinates
$\alpha_{ij}$ and $\beta_{ij}$ $(i , j = 1, 2)$.  We then exponentiate
and write
	$$
	g = \exp \left( X + s \cdot 1_4 \right) \in {\rm PSL}(2|2) \;.
	$$
(As before, $s$ is an arbitrary scalar.)  When the odd variables are
set to zero, the image of $X + s\cdot 1_4$ under $\exp$ lies in $M
\simeq M_{\rm B} \times M_{\rm F}$, since the exponential of a
traceless Hermitian matrix is a positive Hermitian matrix of
determinant one, and the exponential of a traceless skew-Hermitian
matrix lies in ${\rm SU}(2)$.  When the odd variables are included,
exponentiation gives one superdomain of a ${\rm cs}$-manifold with
base $M$.  A second superdomain is defined by repeating the same
procedure with the group unit replaced by the equivalence class of
${\rm diag}(1_2 , - 1_2)$ where $-1_2$ is the ``antipode'' of unity on
${\rm SU}(2) \simeq {\rm S}^3$.  The element playing the role of $X$
is now denoted by $Y$.  To compute the transition functions connecting
$X$ with $Y$, we solve the equation
	$$
	\exp \left( X + s \cdot 1_4 \right) = {\rm diag}(1_2 , -1_2) 
	\exp \left( Y + t \cdot 1_4 \right) \;.
	$$
It is not hard to see that, given the matrix entries of $X$, this
equation on overlapping domains has a unique solution for the matrix
entries of $Y$, and vice versa.  The superdomains of $X$ and $Y$,
together with the transition functions relating them, constitute a
${\rm cs}$-manifold which is the ``backbone'' of the Riemannian
symmetric superspace ${\bf X}_{A_1|A_1}$.

We now describe the supergeometry of ${\bf X}_{A_1|A_1}$ explicitly,
in the superdomain given by $g = \exp (X + s \cdot 1_4)$.  A standard 
result of Lie group theory says that the exponential map pulls the 
Cartan-Maurer form $g^{-1}{\rm d}g$ back to
	$$
	g^{-1}{\rm d}g = {1 - {\rm e}^{-{\rm ad}(X)} \over 
	{\rm ad}(X)} \, {\rm d}X \equiv \sum_{n=0}^\infty
	{ {\rm ad}^n(-X) \over (n+1)! } \, {\rm d}X \;,
	$$
where ${\rm ad}(X)$ acts on ${\rm d}X$ by the commutator: ${\rm ad}(X)
{\rm d}X = [X,{\rm d}X]$, as usual.  The supergeometry is then
expressed by
	$$
	\kappa = \left\langle 
	{1 - {\rm e}^{- {\rm ad}(X)} \over {\rm ad}(X)} \, {\rm d}X \; ,
	\; {1 - {\rm e}^{- {\rm ad}(X)} \over {\rm ad}(X)} \, {\rm d}X 
	\right\rangle \;.
	$$
The Riemannian nature of this geometry can be exhibited more clearly
by introducing a Cartan decomposition.  Let $K \equiv {\rm PSU}(2|2)$
be a {\it compact} real form of the complex Lie supergroup ${\rm
PSL}(2|2) $.  Its elements $k$ act on $X \in {\rm psl}(2|2)$ by
conjugation:
	$$
	X \mapsto kXk^{-1} \equiv {\rm Ad}(k) X \;.
	$$
Utilizing this action, we make a polar decomposition
	$$
	X = {\rm Ad}(k) H \;, \quad H = {\rm diag}(x,-x,iy,-iy) \;,
	$$
where the ``radial'' variables $x$ and $y$ have real range when viewed
as coordinates of the cs-manifold ${\bf X}_{A_1|A_1}$.  (The group
$K$, too, is viewed as a cs-manifold, which is to say we take its
fermions to be complex.)  The differential ${\rm d}X$ decomposes as
${\rm d}X = {\rm Ad}(k) \left( {\rm d}H - {\rm ad}(H) (k^{-1}{\rm d}k)
\right)$, and
	$$
	g^{-1}{\rm d}g = {\rm Ad}(k) \left( {\rm d}H + 
	({\rm e}^{-{\rm ad}(H)} - 1) (k^{-1}{\rm d}k) \right) \;.
	$$
Further evaluation uses the roots $\alpha(H)$ of the adjoint action
${\rm ad}(H)$ on ${\rm psl}(2,2)$.  The roots are called bosonic or
fermionic depending on whether the corresponding eigenvector, or root
vector, is an even or odd element of ${\rm psl}(2,2)$.  Two of the
bosonic roots are zero, and the nonzero ones are $\alpha = \pm 2x$ and
$\alpha = \pm 2iy$, each with multiplicity one.  The fermionic roots
are $\alpha = \pm (x\pm iy)$, and they have multiplicity {\it two}.
(The latter is an exceptional feature possible only in the superworld;
the roots of semisimple Lie {\it algebras} always have multiplicity
one.)  We adopt the convention of giving the multiplicity a sign which
is negative for fermionic roots, and positive for the bosonic ones.
To summarize, a system of positive roots $\alpha$ (with signed
multiplicities $m_\alpha$) is as follows:
	\begin{equation}
	2x \, (+1), \quad 2iy \, (+1), \quad 
	x+iy \, (-2), \quad x-iy \, (-2) \;.
	\label{roots}
	\end{equation}

The metric tensor $\kappa$ is now expressed in terms of polar
coordinates by decomposing $k^{-1}{\rm d}k$ according to root spaces,
	$$
	k^{-1}{\rm d}k = (k^{-1}{\rm d}k)_0 + \sum_{\alpha\not= 0}
	(k^{-1}{\rm d}k)_\alpha \;.
	$$
Insertion into the previous formula for $\kappa$ then gives
	$$
	\kappa = \langle {\rm d}H , {\rm d}H \rangle - 
	4 \sum_{\alpha\not= 0} \sinh^2 {\textstyle{1\over 2}} \alpha(H)
	\left\langle (k^{-1}{\rm d}k)_\alpha \, , (k^{-1}{\rm d}k)_{-\alpha}
	\right\rangle \;.
	$$
The radial term $\langle {\rm d}H , {\rm d}H \rangle = 2 ({\rm d}x^2 +
{\rm d}y^2)$ is obviously nonnegative.  The numerical part of the
other term is nonnegative, too, since $\sinh^2 x \ge 0 \ge \sinh^2
(iy) = - \sin^2 y$ and
	$$
	\left\langle (k^{-1}{\rm d}k)_{2x} \; , (k^{-1}{\rm d}k)_{-2x} 
	\right\rangle \le 0 \le
	\left\langle (k^{-1}{\rm d}k)_{2iy}\; , (k^{-1}{\rm d}k)_{-2iy}
	\right\rangle \;.
	$$
To verify the last inequalities, one needs to use the fact that
$(k^{-1}{\rm d}k)_{\pm 2x}$ and $(k^{-1}{\rm d}k)_{\pm 2iy}$ lie in
sectors where the supertrace acts as $+{\rm Tr}$ and $-{\rm Tr}$,
respectively.

\section{Invariant Berezin integral}
\label{sec:berezin}

For our purposes, an important structure on the symmetric superspace
${\bf X}_{A_1|A_1}$ is its invariant Berezin integral.  Using the
cs-manifold picture, this is described as follows.  It will be
sufficient to use just a single superdomain, say the one centered
around the group unit, which is given by $g = \exp \left( X + s \cdot
1_4 \right)$.  Let $Dg$ denote the invariant Berezin form on ${\bf X}
_{A_1|A_1}$, normalized so that it agrees with the flat Berezin form
	$$
	DX = \prod_{i=1}^3 {\rm d}a_i \, {\rm d}b_i \prod_{j,k=1}^2 
	{\partial^2 \over \partial\alpha_{jk}\partial\beta_{jk}}
	$$
at the group unit.  From the theory of Lie supergroups, we have a
standard formula \cite{berezin} for the Berezinian, or superjacobian,
${\cal B}(X)$ of the exponential map $X \mapsto g = \exp \left( X + s 
\cdot 1_4 \right)$.  It reads
	$$
	{\cal B}(X) = {\rm SDet} \left( {1 - {\rm e}^{-{\rm ad}(X)}
	\over {\rm ad}(X) } \right)\Big|_{{\rm psl}(2,2)} \;.
	$$
We will shortly see that the exponential map has a domain of 
injectivity, ${\cal D}$, delineated by
	$$
	\sqrt{b_1^2 + b_2^2 + b_3^2} < \pi \;,
	$$
while $a_1$, $a_2$ and $a_3$ are unconstrained.  Anticipating this fact, 
we express the invariant Berezin integral in the coordinates $X$ by
	\begin{equation}
	\int_M Dg \, f(g) = \int_{\cal D} DX \, {\cal B}(X)
	\, f\left( \exp(X+s\cdot 1_4) \right) \;.
	\label{invint}
	\end{equation}
If $f$ vanishes fast enough at infinity on ${\rm H}^3$, this
expression is correct as it stands, and there is no need to add any
boundary distributions (which often appear in coordinate expressions
for superintegrals) to the right-hand side.

To establish the last claim, we make the Berezinian ${\cal B}(X)$ more
explicit by introducing polar coordinates $X = {\rm Ad}(k) H$, as
before.  The Berezinian is a radial function,
	$$
	{\cal B}(X) = {\cal B}({\rm Ad}(k) H) = {\rm SDet}\, {\rm Ad}(k)
	\big|_{{\rm psl}(2,2)} \times {\cal B}(H) = {\cal B}(H) \;,
	$$
by the multiplicativity of ${\rm SDet}$ and the unimodularity of ${\rm
PSL}(2|2)$.  Recall our convention of taking the multiplicity $m_\alpha$ 
to be negative (positive) for a fermionic (bosonic) root $\alpha$.  If 
$\Delta^+$ is a system of positive roots as specified in (\ref{roots}), 
the Berezinian can be written in the form
	$$
	{\cal B}(H) = \prod_{\alpha\in\Delta^+} \left(
	{\sinh {1\over 2}\alpha(H) \over {1\over 2}\alpha(H)}
	\right)^{2m_\alpha} \;.
	$$
The existence of the bosonic roots $\pm 2iy$ makes ${\cal B}(H)$
vanish for $y \in \pi{\mathbb Z}\setminus \{0\}$.  Hence the
injectivity domain of the exponential map is given by $|y| < \pi$.
This translates into the condition $\sqrt{b_1^2 + b_2^2 + b_3^2} <
\pi$ stated earlier.  Moreover, the singularities of ${\cal B}(H)$
caused by the vanishing of the fermionic roots are located at $x = 0$,
$y \in 2\pi{\mathbb Z} \setminus \{ 0 \}$.  Therefore ${\cal B}(H)$ is
analytic and regular for $x \in {\mathbb R}$ and $- \pi < y < \pi$.
By extension, ${\cal B}(X)$ is analytic and regular on the injectivity
domain ${\cal D}$, and by the vanishing of ${\cal B}(X)$ on the
boundary $\partial{\cal D}$, no boundary distributions can occur in
the coordinate expression (\ref{invint}) of the invariant Berezin
integral.

We now turn to a property of the Berezin integral which is a
prerequisite for understanding the field theory partition function.
Consider the normalization integral $\int_{\cal D} DX \, {\cal B}(X)$.
As it stands, this makes no sense, for the presence of the hyperbolic
sector ${\rm H}^3$ renders the integration domain ${\cal D}$ infinite.
To fix the problem, we insert some convergence factor ${\rm e}^ {-
\epsilon h(\exp(X+s\cdot 1_4))}$, with $\epsilon$ small, and define 
the normalization integral by
	$$
	N(\epsilon) = \int_{\cal D} DX \, {\cal B}(X)
	\, {\rm e}^{-\epsilon h(\exp(X+s\cdot 1_4))} \;.
	$$
The question now is how to choose the function $h$.  We want $h$ to be
effective as a regulator but, at the same time, it should preserve as
many symmetries as possible.  This leads us to choose $h$ as a radial
function: $h(g) = h \left( {\rm Ad}(k) \exp(H+s\cdot 1_4) \right) = 
h \left( \exp(H+s\cdot 1_4) \right)$.  The first guess is to take
	\begin{eqnarray*}
	{\rm STr}\,{\rm Ad}(g) &=& {\rm STr}\,{\rm e}^{{\rm ad}(H)} \\
	&=& 1 + 1 + {\rm e}^{2x} + {\rm e}^{-2x} + {\rm e}^{2iy} 
	+ {\rm e}^{-2iy} \\
	&&- 2\,{\rm e}^{x+iy} - 2\,{\rm e}^{x-iy} - 2\,{\rm e}^{-x+iy}
	-2\,{\rm e}^{-x-iy} \\	
	&=& - 2 + 4 (\cosh x - \cos y)^2 \;.
	\end{eqnarray*}
Near $x = y = 0$ this varies as a fourth power $(x^2 + y^2)^2$, which
has the undesirable feature that the Hessian degenerates to zero at
that point.  (This degeneracy is an immediate consequence of the
vanishing of the Killing form on ${\rm psl}(2,2)$: ${\rm STr} \, {\rm 
ad}^2(X) = 0$.)  A better choice is
	$$
	h(g) = \sqrt{2+{\rm STr}\,{\rm Ad}(g)} = 2(\cosh x - \cos y) \;.
	$$
The right-hand side of this equation shows that the square root exists
as an {\it analytic} function on all of the ${\rm cs}$-manifold ${\bf
X}_{A_1|A_1}$.  For future use, note that the Taylor expansion of $h$
around the identity coset reads
	$$
	h \left( \exp(X + s\cdot 1_4) \right) = 
	{\textstyle{1\over 2}} \langle X , X \rangle + ... \;.
	$$

For the choice of regulator ${\rm e}^{-\epsilon h}$ made, we now
evaluate the normalization integral $N(\epsilon)$.  This can be done
by a localization principle called the Parisi-Sourlas-Efetov-Wegner
theorem in disordered electron physics.  Its most recent version has 
been stated and proved by Schwarz and Zaboronsky \cite{sz}.  

In the specific setting at hand, the localization principle is
formulated as follows.  Let ${\bf X}$ be a Riemannian symmetric
superspace with symmetry group $G$ and invariant Berezin form $D{\bf
x}$.  Pick an odd generator ${\cal F}$ of $G$, and denote by $\Xi$ the
Killing vector field representing the action of ${\cal F}$ on ${\bf
X}$.  Then, if $f$ is a function invariant under $\Xi$ ({\it i.e.},
$\Xi f = 0$), the integral $\int D{\bf x}\,f$ {\it localizes on the
zero locus} of the vector field $\Xi$ (viewed as a differential
operator).  This means that the integral is determined by the values
of the integrand, and a finite number of its derivatives, at the zero
locus of $\Xi$.  If $\Xi$ is expressed in even and odd local
coordinates $x^1,...,x^p$ and $\xi^1, ..., \xi^q$ by $\Xi = a^i(x,\xi)
\partial/\partial\xi^i + \alpha^i(x,\xi) \partial / \partial x^i$, the
zero locus of $\Xi$ is defined as the set of solutions of the
equations $a^i(x,0) = 0$ $(i = 1, ..., q)$.

The mechanism behind this version of the localization principle can be
stated in a few sentences.  Let the zero locus of $\Xi$ be denoted
by $R_\Xi$ and its complement by ${\cal C}$.  On the latter, the
supergroup generated by ${\cal F}$ acts freely, or without fixed
points.  This allows us to introduce local coordinates on ${\cal C}$ 
such that the Killing vector field $\Xi$ takes the simple form $\Xi =
\partial / \partial \xi^1$.  Now, by $\Xi$-invariance neither the 
function $f$ nor the Berezin form $D{\bf x}$ carries any dependence on
$\xi^1$.  Doing the Fermi integral over $\xi^1$ therefore yields zero.
This reasoning breaks down on the set $R_\Xi$, but it remains valid
outside an arbitrarily small neighborhood of $R_\Xi$.  As a result,
the integral $\int D{\bf x} \,f$ can only depend on a finite number of
terms in the Taylor expansion of $f$ at $R_\Xi$.

This principle allows us to compute the normalization integral $N
(\epsilon)$ rather easily.  Because the regulator ${\rm e}^ {-\epsilon
h}$ is a function with a high degree of symmetry, there exist several
odd vector fields with the required properties in order for the
localization principle to take effect.  It will be sufficient to use
just one of them.  Fix $i,j \in \{1,2\}$, and let $E_{ij}$ be the $2
\times 2$ matrix whose entries are zero everywhere except on the
intersection of the $i$-th row with the $j$-th column where the entry
is unity. Put
	$$
	{\cal F} = \pmatrix{0 &E_{ij}\cr 0 &0\cr} \;,
	$$
and, with $\sigma$ an odd parameter, define $\Xi$ by
	$$
	(\Xi f)(g) = {\partial\over\partial\sigma} f \left( 
	{\rm e}^{\sigma{\cal F}} g\,{\rm e}^{-\sigma{\cal F}}\right) \;.
	$$
$\Xi$ is a Killing vector field since conjugation by ${\rm e}^{
\sigma {\cal F}}$ preserves the supergeometry of ${\bf X}_{A_1|A_1}$. 
Moreover, any radial function $f$ satisfies $f \left( {\rm e}^{\sigma
{\cal F}} g\,{\rm e}^{-\sigma{\cal F}} \right) = f(g)$, and is
therefore invariant w.r.t.~$\Xi$.  In particular, the regulator ${\rm
e}^{-\epsilon h}$ is $\Xi$-invariant.  Hence, by the principle stated,
the normalization integral $N(\epsilon)$ localizes on the zero locus
of $\Xi$.  We claim that the latter consists of only a single point, 
namely the origin $X = 0$ of ${\bf X}_{A_1|A_1}$.  To see that this is 
so, we examine the infinitesimal action of ${\rm e}^{\sigma{\cal F}}$ on
the radial space, or the Cartan subalgebra.  This action is determined
by the root corresponding to the root vector ${\cal F}$, which is one
of the set $\pm(x\pm iy)$ and vanishes only for $x=y=0$, or $X=0$, as
claimed.  (Note that the Riemannian nature of ${\bf X}_{A_1|A_1}$ is
crucial for this argument.  If we were working on ${\rm PSL}_{\mathbb
R}(2|2)$ or ${\rm PSU}(2|2)$, the roots would be $\pm(x\pm y)$ or $\pm 
i(x\pm y)$, giving a much bigger set for the zero locus of $\Xi$.)
The implication is that we may Taylor expand the integrand around 
$X = 0$ and reduce $N(\epsilon)$ to a Gaussian integral: 
	$$
	N(\epsilon) = \int DX\,\exp - {\epsilon\over 2}
	\langle X , X \rangle \;,
	$$
which is readily calculated to be $N(\epsilon) = \pi^3 \epsilon$.
Here we recognize the utility of defining the regulator $h$ by taking
a square root, so as to have a nondegenerate Hessian at $X = 0$: a
vanishing Hessian would have forced us to expand to fourth order in
$X$.

We thus see that the normalization integral $N(\epsilon)$ is
completely determined by Gaussian fluctuations around the point $X =
0$.  The same is true for a large class of integrals on ${\bf X}_{
A_1|A_1}$.  Note that the symmetry group of $X_{A_1|A_1}$ provides
us with a total of 8 fermionic Killing vectors.  In order for
localization onto $X = 0$ to occur, it is sufficient for the
integrand to be invariant under a {\it single one} of these vector
fields.  (Recall that this is a consequence of the fermionic root
system being $\pm(x\pm iy)$.)  The quantum numbers $i,j = 1,2$
counting these symmetries originate from the distinction between
retarded and advanced Green functions of the disordered electron
system.  Let us put
	\begin{equation}
	{\cal F}_{ij}^+ = \pmatrix{0 &E_{ij}\cr 0 &0\cr}, \quad
	{\cal F}_{ij}^- = \pmatrix{0 &0\cr E_{ji} &0\cr}
	\label{susycharges}
	\end{equation}
and call the sector $i = 1$ ``retarded'' and $i = 2$ ``advanced''.
The correlation functions or observables which involve only retarded
Green functions will be invariant under the Killing vector fields of
${\cal F}_{22}^\pm$, and those involving only advanced information
will be invariant under the Killing vector fields of ${\cal F}_{11}
^\pm$.  As a consequence of the localization principle, such observables
are trivial.  We will return to this point later, in the field-theoretic
setting.

We can get a more complete perspective on the Berezin integral for 
${\bf X}_{A_1|A_1}$ by making the polar decomposition
	$$
	X = {\rm Ad}(k) H \;.
	$$
Under this substitution, the flat Berezin form $DX$ transforms as
	$$
	DX \longrightarrow j(H) dH \, Dk \;.
	$$
Here $Dk$ is an invariant Berezin form on $K / T$, where $T \simeq
{\rm U}(1) \times {\rm U}(1)$ is a maximal torus, $dH = {\rm d}x \,
{\rm d}y$ is a Euclidian radial measure, and $j(H) = {\rm SDet} \,
{\rm ad}(H)\big|_{{\cal T}_o(K/T)}$ is the superdeterminant of ${\rm
ad}(H)$ acting on the tangent space of $K / T$ at the origin $o \equiv
T$. The product of the Berezinians of the two transformations $(k,H)
\mapsto X = {\rm Ad}(k)H$ and $X \mapsto g = \exp \left( X + s \cdot 
1_4 \right)$ is
	$$
	J(H) = j(H) {\cal B}(H) = {\sinh^2 x \sin^2 y \over
	(\cosh x - \cos y)^4 } \;.
	$$
The invariant Berezin integral for $X_{A_1|A_1}$ now takes the form
	\begin{equation}
	\int_M Dg \, f(g) = {\cal R}[f] + \int \left( \int_{K/T} Dk\, 
	f\left( k \exp(H+s\cdot 1_4) k^{-1} \right) \right) J(H) dH
	\label{polar}
	\end{equation}
where the radial integal is restriced to run over a Weyl chamber $0 <
x < \infty$ and $0 < y < \pi$.  The first term on the right-hand side
is a boundary distribution, which shows up as a consequence of the
singularity of the function $J(H)$ at the origin $x = y = 0$.  The
general theory of boundary distributions for polar coordinate
integrals on symmetric superspaces yields an explicit formula for
${\cal R}[f]$.  We will skip the details here, as they form the
subject of a separate paper \cite{bz}.  In brief, the idea is to vary
the ``bulk term'' (the second term on the right-hand side of
(\ref{polar})) by an infinitesimal isometry of the superspace.  By
partial integration on $K / T$, this variation can be manipulated to
become the radial integral of an exact form, and application of
Stokes' formula then converts it into an integral over the boundary of
the radial space.  By invariance of the complete integral, this
boundary term must be exactly cancelled by the variation of ${\cal
R}[f]$.  In this way, one obtains the result
	$$
	{\cal R}[f] = (L f)(g_0)\;,
	$$
where $L$ denotes the Laplace-Beltrami operator, and $g_0$
the identity coset ${\rm e}^s \cdot 1_4$. 

The formula (\ref{polar}) holds for any integrable function $f$,
radial or not.  We now apply it to the normalization integral
$N(\epsilon)$.  In that case the second term on the right-hand side of
(\ref{polar}) disappears, since $f = {\rm e}^{-\epsilon h}$ is radial
and $\int Dk = 0$.  The last statement follows from the fact that the
volume of $K = {\rm PSU}(2|2)$ is zero, as is the volume of ${\rm U}
(2|2)$.  It can also be understood from the localization principle, by
noting that $K$ acts on $K/T$ without fixed points.  Hence,
	$$
	N(\epsilon) = \int_M Dg \, {\rm e}^{-\epsilon h(g)} = 
	{\cal R}[{\rm e}^{-\epsilon h}] = \left(L \, {\rm e}^
	{-\epsilon h} \right) (g_0) \;,
	$$
and by applying the Laplacian to ${\rm e}^{-\epsilon h}$ at the 
identity coset $g_0$, we recover the result $N(\epsilon) = \pi^3 
\epsilon$. 

\section{The $A_1|A_1$ nonlinear sigma model}
\label{sec:cft}

After this spacious presentation of mathematical background, we are
going to discuss a nonlinear sigma model of maps from $\Sigma$, the
configuration space for a single electron of the two-dimensional
electron gas, into the Riemannian symmetric space ${\bf X}_{A_1|A_1}$.
Although the latter is not a group, its complexification is.  To
remind ourselves of this fact, we denote the field by $g$.  It is also
important to keep in mind that the target space does not have a
representation in ${\rm Mat}(2|2)$ but consists of {\it equivalence
classes} of supermatrices.  This places a constraint on the terms that
may appear in the field-theory Lagrangian.

The principal term of the nonlinear sigma model action is given by
	$$
	S_0 = {1 \over 2\pi} \int_\Sigma d^2 x \, \left\langle g^{-1}
	\partial g , g^{-1} \bar\partial g \right\rangle \;.
	$$
To break parity of the field theory (or parity of the disordered 
electron gas) we need a term of the Wess-Zumino-Novikov-Witten type:
	$$
	\Gamma = {1\over 24\pi} \int_\Sigma {\rm d}^{-1} \left\langle 
	g^{-1}{\rm d}g,[g^{-1}{\rm d}g,g^{-1}{\rm d}g]\right\rangle \;,
	$$
Because of the presence of $M_{\rm F} \simeq {\rm S}^3$ in the base
$M = M_{\rm B} \times M_{\rm F}$ of ${\bf X}_{A_1|A_1}$, single-valuedness
of ${\rm e}^{-ik\Gamma}$ quantizes the coupling $k$ to be an integer
\cite{novikov}.  The two functionals $S_0$ and $\Gamma$ have zero
modes (the constant fields), so we include a term proportional to
	$$
	S_{\rm reg} = \int_\Sigma d^2x \, h(g)
	$$
to regularize them.  The complete action is then
	\begin{equation}
	S = f^{-2} S_0 + ik\Gamma + \epsilon S_{\rm reg} \;,
	\label{action}
	\end{equation}
where $f$ is a coupling constant, and $\epsilon$ is a positive
infinitesimal.  We call this two-parameter field theory the $A_1|A_1$
nonlinear sigma model with Wess-Zumino term.  The theory does not have
the invariance under $g(z,\bar z) \to \Omega(z) g(z,\bar z) \bar
\Omega (\bar z)$ characteristic of a WZW model, unless $f^2=1/|k|$.

Note that the theory (for $\epsilon = 0$) is invariant under
infinitesimal chiral transformations $g(x) \mapsto X_{\rm L} g(x) -
g(x) X_{\rm R}$ with $X_{\rm L} , X_{\rm R} \in {\rm psl}(2|2)$.  This
exponentiates to an invariance under
	$$
	g(x)\mapsto g_{\rm L}^{\vphantom{-1}}\,g(x)\,g_{\rm R}^{-1}
	$$
as long as $g_{\rm L}, g_{\rm R} \in {\rm PSL}(2|2)$ are close enough
to the group unit.  It has to be said that such a transformation is
not an isometry of ${\bf X}_{A_1|A_1}$ for general $g_{\rm L}, g_{\rm
R}$.  (In particular, recall that the left and right actions of ${\rm 
SL}(2,{\mathbb C})$ on ${\rm H}^3 \subset {\bf X}_{A_1|A_1}$ are related 
to each other by a unitarity condition.)  Nevertheless, the functional 
integral does remain invariant, by a functional generalization of what 
is called Cauchy's theorem in complex analysis.  The situation is the 
same as for the Haar integral of a compact Lie group $U$:
	$$
	\int_U F(g) dg = \int_U F(g_{\rm L}^{\vphantom{-1}} 
	\, g \, g_{\rm R}^{-1}) dg \;,
	$$
the invariance of which is not restricted to $g_{\rm L}, g_{\rm R} \in
U$, but can hold (depending on the analytic properties of $F$) more
generally for $g_{\rm L}$ and $g_{\rm R}$ in the complexification of
$U$.

The invariance under $G_{\rm L} \times G_{\rm R}$ is the chiral
symmetry which we expect to emerge for the disordered electron system
at criticality, in view of our discussion of the loop-group coherent
state path integral for the superspin chain in Section \ref{sec:spin}.
Away from the critical point, where chiral symmetry is broken to the
diagonal $(g_L = g_R)$, the low-energy physics should be described by
Pruisken's nonlinear sigma model.  Hence, Pruisken's model has to sit
inside the $A_1|A_1$ model, and the field $Q$ of the former is somehow
related to the field $g$ of the latter.  To describe the relation,
write $G \equiv {\rm PSL}(2|2)$, and let $H \subset G$ be the subgroup
of elements $h$ that stabilize $\Sigma_3 = {\rm diag}(+1,-1,+1,-1)$
under conjugation: $h \Sigma_3 h^{-1}$.  (This is the same $\Sigma_3$
as in Section \ref{sec:nlsm}, but we have found it convenient to
rearrange the basis, and thus the ordering of matrix entries, from
``retarded first, advanced last'' to ``bosons first, fermions last''.)
The quotient $G / H$ is naturally isomorphic to ${\rm GL}(2|2) / {\rm
GL}(1|1) \times {\rm GL}(1|1)$.  We can implement the projection from
$G$ to $G / H$ by mapping $g$ onto $g \Sigma_3 g^{-1}$, and this
descends to a projection from ${\bf X}_{A_1|A_1}$ onto ${\rm U}(1,1|2) /
{\rm U}(1|1) \times {\rm U}(1|1)$, which is Pruisken's target space.
We are thus led to put
	\begin{equation}
	Q = {\rm Ad}(g) \Sigma_3 = g \, \Sigma_3 \, g^{-1} \;.  
	\label{Q}
	\end{equation}
There exists another strong motivation for making this identification.
Requirement 7 of our list says that the density of states of the
disordered electron system is {\it noncritical} at the critical point.
It must therefore be represented in the field theory by an operator of
vanishing scaling dimension.  In Pruisken's theory, the density of
states is known to be represented by the operator ${\rm STr} \Sigma_3
Q$.  To arrange for its scaling dimension to be zero, we should
construct it from the adjoint representation of ${\rm PSL}(2|2)$,
which has vanishing quadratic Casimir.  (From a one-loop computation, 
the scaling dimension of a local operator transforming according to a
representation $\rho$, is proportional to the quadratic Casimir
invariant evaluated on $\rho$.  It has been conjectured \cite{bzv}
that in the present case this relation is not changed by higher-loop 
corrections.) This observation, together with the necessity to 
project out $H$, strongly suggests $Q = {\rm Ad}(g)\Sigma_3$.

For the purpose of calculating $\Omega(a_0,a_1;b_0,b_1)$, the 
correlator of spectral determinants of Section \ref{sec:cc}, we 
perturb the field theory by adding an extra term
	$$
	S_\omega = \Lambda^2 \int_\Sigma d^2x \, {\rm STr} \left(
	\omega {\rm Ad}(g) \Sigma_3 - \omega \Sigma_3 \right)
	$$
where, in the ordering chosen (bosons first, fermions last),
	$$
	\omega = {\rm diag} (\ln a_0,-\ln b_0, \ln a_1,-\ln b_1) \;,
	$$
and $\Lambda$ is a UV cutoff.  This form of $S_\omega$ follows from 
the frequency term $L_\omega$ of Pruisken's model, on setting $Q = 
{\rm Ad}(g) \Sigma_3$ and reordering the matrix entries.

Let us then restate the basic proposition of the present paper: the
$A_1|A_1$ nonlinear sigma model as defined above (with coupling
constants $f$ and $k$ that will be fixed in due course) is the
conformal field theory describing the critical physics of electron
delocalization at the transition between quantum Hall plateaus.

Our job in the following is to check the list of requirements laid
down in Section \ref{sec:constraints}.  The first one was that the
functional integral be well-defined and stable.  We have been careful
to define the target space as a symmetric superspace with a Riemannian
metric on its base, so that the numerical part of $S_0$ is nonnegative
and becomes zero only on the constant fields.  In the absence of a WZW
term, this would already guarantee the stability of the functional
integral $\int {\rm e}^{-S}$.  However, with a WZW term present, the
situation is less benign.  Although $ik\Gamma$ takes imaginary values
in the Fermion-Fermion (FF) sector (or when the target is a compact
group), it is {\it real-valued} in the Boson-Boson (BB) sector.
Indeed, the 3-linear Lie algebra form underlying the WZW term,
	$$
	\Omega(X,Y,Z) = i \big\langle X , [ Y , Z ] \big\rangle
	= i {\rm Tr} \, X [ Y , Z ] \;,
	$$
is imaginary for $X, Y , Z \in {\rm su}(2)$ but is real when $X$, $Y$,
and $Z$ lie in $i{\rm su}(2)$, the BB-part of the tangent space at the
origin of ${\bf X}_{A_1|A_1}$.  The WZW term can have either sign, and
therefore jeopardizes the existence of the functional integral, unless
it is bounded by $|{\rm Re}(ik\Gamma)| \le S_0 / f^2$.  We are now
going to show that this bound is obeyed if and only if
	\begin{equation}
	1 / f^2 \ge |k| \;.
	\label{stability}
	\end{equation}
For that purpose, we denote the field in the BB-sector by $M = g_{\rm BB}$,
and parametrize it by hyperbolic polar coordinates with range $0 \le \psi$, 
$0 \le \theta \le \pi$, and $0 \le \phi \le 2\pi$:
	$$
	M = \pmatrix{\cosh\psi + \sinh\psi \cos\theta 
	& \sinh\psi \sin\theta \, {\rm e}^{i\phi} \cr
	\sinh\psi \sin\theta \, {\rm e}^{-i\phi}
	& \cosh\psi - \sinh\psi \cos\theta \cr} \;.
	$$
In these coordinates, the BB-part of $S_0$ is given by
	$$
	{1\over 2} \int d^2x \, {\rm Tr} \left( M^{-1} \partial_\mu M 
	\right)^2 = \int d^2x \Big( (\partial_\mu \psi)^2 + \sinh^2 \psi 
	\, \left( (\partial_\mu \theta)^2 + \sin^2 \theta \, (\partial_\mu 
	\phi)^2 \right) \Big) \;,
	$$
and the WZW 3-form is expressed by
	\begin{eqnarray*}
	{i \over 3} {\rm Tr} \left( M^{-1} {\rm d}M \right)^{\wedge 3}
	&=& 4 \sinh^2 \psi \, {\rm d}\psi \wedge \sin\theta \, 
	{\rm d}\theta \wedge {\rm d}\phi \\
	&=& {\rm d} \left( \sinh(2\psi) - 2\psi \right) \wedge
	\sin\theta \, {\rm d}\theta \wedge {\rm d}\phi \;.
	\end{eqnarray*}
Now, the following two inequalities,
	$$
	\Big| \epsilon_{\mu\nu} \sin\theta\, \partial_\mu \theta \, \partial_\nu
	\phi \Big| \le {\textstyle{1 \over 2}} \left( (\partial_\mu \theta)^2
	+ \sin^2 \theta \, (\partial_\mu \phi)^2 \right) \;,
	$$
and
	$$
	{\textstyle{1 \over 2}} \sinh (2\psi) - \psi \le \sinh^2 \psi \;,
	$$
are immediate from elementary considerations.  Using them, we estimate
	\begin{eqnarray*}
	\Big| {\rm Re} (ik\Gamma) \Big| = \Big| {k\epsilon_{\mu\nu} \over
	8\pi} \int_\Sigma d^2x \left( \sinh(2\psi) - 2\psi \right) 
	\sin\theta \, \partial_\mu \theta \, \partial_\nu \phi \Big| \\
	\le {|k|\over 8\pi} \int_\Sigma d^2x \, \sinh^2 \psi \left( 
	(\partial_\mu \theta)^2 + \sin^2 \theta \, (\partial_\mu \phi)^2 
	\right) \le |k| S_0 \;.
	\end{eqnarray*}
The last expression is obviously bounded from above by $S_0 / f^2$ if
$|k| \le 1/f^2$.  Moreover, this bound is optimal as can be seen from
taking $\partial_\mu \psi = 0$ and sending $\psi \to \infty$.  Hence,
the condition (\ref{stability}) is necessary for stability of the
functional integral. 

We should mention that the same condition appeared in \cite{bvw}.
There, it arose as a consequence of the reality of the Ramond-Ramond
and Neveu-Schwarz fluxes of the fivebranes that determine the ${\rm
AdS}_3 \times {\rm S}^3$ background (the target space of the nonlinear
sigma model) and the values of the couplings $1/f^2$ and $k$.

\section{BRST invariance}
\label{sec:brst}

The second requirement of our list is manifest conformal invariance of
the field theory.  This is not a pressing issue and is postponed until
the next section.  The most urgent problem to address are requirements
3 and 4, concerning the normalization of the partition function of the
field theory and the triviality of some of its correlation functions.
These two points are closely related (the former is a consequence of
the latter) and can be dealt with in a single shot. 

Consider the correlator of spectral determinants $\Omega(a_0, a_1 ;
b_0 , b_1)$ defined in (\ref{speccorr}).  For elementary reasons that
were spelled out in Section \ref{sec:cc}, it reduces to unity on
setting either $a_0 = a_1$ or $b_0 = b_1$.  How does the field theory
manage to perform the trick of reproducing this feature?  The short
answer is that the theory becomes {\it topological} and undergoes
dimensional reduction, by one or several of its many supersymmetries.
(The mechanism is nicely explained in \cite{brst}.)  To see concretely
how the reduction works, we consider the case $a_0 = a_1$.  For these
parameter values, the action functional $S + S_\omega$ acquires a
global invariance under supersymmetry transformations in the retarded
sector ($i=j=1$), since the first and third entry of the diagonal
matrix $\omega$ are now equal.  The transformations are given by
	$$
	\delta g(x) = [ \sigma {\cal F}_{11}^\pm\,,\,g(x) ] \;,
	$$
where $\sigma$ is an odd parameter, and ${\cal F}_{11}^\pm$ are two of
the eight fermionic generators ${\cal F}_{ij}^\pm$ of ${\rm psl}
(2,2)$, see (\ref{susycharges}).  Viewing these generators as BRST
operators, we may express the global invariance of $S + S_\omega$ by
saying that the theory is BRST-closed.  The BRST transformations can
be recast in the form
	$$
	g^{-1} \delta g = \left({\rm Ad}(g)-1\right)
	\sigma {\cal F}_{11}^\pm \;.
	$$
To exploit the BRST symmetry, we note that the only fixed point of the
BRST transformations is the constant field configuration $g(x) \equiv
g_0 \equiv {\rm e}^s \cdot 1_4$.  The reason is the same as in the
zero-dimensional case: ${\rm Ad}(g)$ on any one of the fermionic root
vectors ${\cal F}_{ij}^\pm$ equals unity if and only if $g$ is the
{\it origin} of ${\bf X}_ {A_1 |A_1}$.  (Once again, the Riemannian
nature of ${\bf X}_ {A_1|A_1}$ is crucial here, and the situation
would be less favorable if we were working on ${\rm PSL}_{\mathbb R}
(2|2)$ or ${\rm PSU}(2|2)$.)  Being fermionic, the generators ${\cal
F}_{11}^\pm$ square to zero. As a result, if we {\it exclude} from the
functional integral an infinitesimal ball $B_\varepsilon$ surrounding
the BRST fixed point $g(x) = g_0$, the theory becomes BRST-exact.  In
other words, on the complement of $B_\varepsilon$ the functional
integrand is a ``total derivative'' w.r.t.~some odd collective
coordinate $\sigma$.  Doing the Fermi integral over $\sigma$ yields
zero.

The field theory then collapses onto $B_\varepsilon$, and we can
compute the functional integral by simply carrying out leading-order
perturbation theory around the configuration $g(x) = g_0$.  The rest
is easy.  Expansion of $S + S_\omega$ around that configuration up to
quadratic order in $X$ gives
	$$
	S^{(2)} = \int d^2 x \left( {1 \over 2\pi f^2} 
	\langle \partial X , \bar\partial X \rangle + {\epsilon\over 2} 
	\langle X , X \rangle + {\Lambda^2 \over 2} \langle \omega , 
	{\rm ad}^2(X) \Sigma_3 \rangle \right) \;,
	$$
and by doing the Gaussian integral over $X$ we obtain the expression
	\begin{equation}
	{\rm const} \times {\rm Det}(K) \,
	{{\rm Det}\left( K + \Lambda^2 \ln(a_1 b_0) \right)	
	\,{\rm Det}\left( K + \Lambda^2 \ln(a_0 b_1) \right)\over
	{\rm Det}\left( K + \Lambda^2 \ln(a_0 b_0) \right)	
	\,{\rm Det}\left( K + \Lambda^2 \ln(a_1 b_1) \right)} \;,
	\label{dets}
	\end{equation}
where $K$ is the operator $K = - (\pi f^2)^{-1} \partial\bar\partial 
+ \epsilon$.  The first factor is a constant due to the choice of
normalization of the functional integral measure, which we are still
free to specify.  It is understood that $a_0 = a_1$, as was assumed at the
outset of our calculation.  Then the determinants in the numerator and
denominator cancel pairwise, and we are left with
	$$
	\Omega(a,a;b_0,b_1) = {\rm const} \times
	{\rm Det}(K) \;.
	$$
Obviously, in order for this to be unity, we must choose the normalization
to be 
	$$
	{\rm const} = {\rm Det}^{-1}(K) \;.
	$$
We can make this choice explicit by shifting the action $S \to S + \ln 
{\rm Det}(K)$.  Alternatively, we can introduce a ghost field, say a 
complex free boson $\varphi$ (or, equivalently, two real free bosons)
and pass to
	$$
	S^\prime = S+\int_\Sigma d^2x\,\bar\varphi(x)(K\varphi)(x) \;.
	$$
What we did for $a_0 = a_1$ can be repeated {\it mutandis verbis} for 
$b_0 = b_1$.  Summarizing the two cases we have
	$$
	\Omega(a,a;b_0,b_1) = 
	\Omega(a_0,a_1;b,b) = 1 \;.
	$$
In particular, it follows that the partition function is normalized 
to unity: $Z = \Omega(1,1;1,1) = 1$.

One might now object that we are enforcing normalization by appealing
to a ghost field $\varphi$, the origin of which we did not explain.
This objection has to be taken seriously.  Normalization of the
partition function is a very basic and robust feature of the
supersymmetric field-theory formalism for disordered electron systems.
It holds independently of all the parameters of the system (such as
disorder strength, system size, lattice constant {\it etc.}) and in
particular, applies whether the system is critical or not.  And,
indeed, in Pruisken's nonlinear sigma model the normalization always
{\it is} unity.  This happens by the same BRST mechanism that was
utilized above, except that the bosonic and fermionic degrees of
freedom are now equal in number, so all determinants cancel exactly.
Thus, normalization is automatic, and no fine-tuning of the parameters
is needed.

One might therefore be worried by the dependence of the normalization
factor ${\rm Det}^{-1}(K)$ on the coupling $f$ and the ultraviolet
cutoff $\Lambda$.  (The dependence on $\epsilon$ is {\it not} an
issue.  This parameter is a positive infinitesimal whose only role is
to regularize the zero modes.  Alternatively, we could set $\epsilon =
0$ and modify the field configuration space by taking the quotient by
the zero modes.)  Our response is this.  The theory we have written
down is not claimed to be applicable to the whole range of parameters
of the quantum Hall system, but is specifically made for its critical
point.  The theory is imagined to be the endpoint of a renormalization
group (RG) trajectory starting, say, from Pruisken's model at weak
coupling (and $\sigma_{xy} = 1/2$).  Our working hypothesis is that
the ${\rm boson} | {\rm fermion}$ count, which is $4|4$ in Pruisken's
model is promoted to $6|8$ in the $A_1|A_1$ model at the fixed point.
In Pruisken's model the normalization is unity, and we can be sure
that supersymmetry acts to keep it there all the way along the RG
trajectory.  Therefore two real bosonic ghosts must appear to absorb
the change in normalization due to the mismatch 6 versus 8.  From this
perspective, the quantities $f$ and $\Lambda$ entering ${\rm Det}(K)$
are not parameters at our disposal, but are numbers determined by the
RG flow.

If the theory is to make sense as an RG fixed point for the disordered
electron system, what we must demand is that changing the cutoff
$\Lambda$ preserves the normalization.  And, in fact, it does: the
operator associated to the coupling constant $f$ is truly marginal in
the bosonic ghost theory and also in the $A_1|A_1$ model (see the next
section).  Therefore, once the normalization has been set to unity, it
stays there when a RG transformation is applied.

Here we should mention that our way of introducing a bosonic ghost is
not unique.  We have opted for the simplest choice, which is to take
$\varphi$ to be a free field.  In contrast, in the string-motivated
construction of \cite{bvw}, there appears a similar ghost which
couples to an expression quadratic in the left-invariant fermionic
currents of the field $g$.  This coupling is apparently required for
the consistency of the nonlinear sigma model as a string theory (where
the ghost not only serves to cancel the conformal anomaly, but is also
needed for $N = 2$ worldsheet superconformal invariance).  However,
there apparently exists no motivation for such a coupling from the
perspective of the disordered electron gas, so we will not consider
it.
 
The above mechanism of dimensional reduction would not be convincing
if it were limited to the partition function and the spectral
correlator $\Omega$. Fortunately, it applies more generally, and the
same reduction occurs for any observable which is already present in
Pruisken's theory and is invariant under supersymmetry (or BRST)
transformations in the retarded or advanced sector.  The argument
goes as follows.  Consider any correlation function
	\begin{equation}
	\Big\langle F_1(Q(x_1))F_2(Q(x_2))\dots F_p(Q(x_p))\Big\rangle 
	\;, \label{cf}
	\end{equation}
where the $F_i$ are local functionals of $Q$.  Now assume BRST
invariance in the retarded or advanced sector.  Then, by the same
reasoning as before, the functional integral will localize onto the
BRST fixed point $g(x) \equiv g_0$.  What remains to be done is the 
(Gaussian) integral over the small fluctuations around $g_0$.  For 
this purpose we decompose the Lie algebra of $G = {\rm PSL}(2|2)$ as
	$$
	{\rm Lie}(G) = {\rm Lie}(H) + {\cal P} \;,
	$$
where $H \subset G$ still is the stabilizer of $\Sigma_3 \in {\rm 
Lie}(G)$ under the adjoint action.  We then put
	$$
	g(x) = {\rm e}^{Y(x)} {\rm e}^{X(x)} g_0 \;,
	$$
where $X(x) \in {\rm Lie}(H)$ and $Y(x) \in {\cal P}$.  By design,
$Q$ depends only on $Y$:
	$$
	Q(x) \Sigma_3 = {\rm e}^{Y(x)} \, \Sigma_3 \, {\rm e}^{-Y(x)} 
	\Sigma_3 = {\rm e}^{2 Y(x)}\;.
	$$
Substituting these expressions into the functional integral and
expanding around $g_0$, we find that the integrand separates into two
factors, one depending on $X$ and another depending on $Y$:
	$$
	\big\langle F_1(Q(x_1)) \dots F_p(Q(x_p)) \big\rangle = 
	\big\langle 1 \big\rangle_X \times \big\langle 
	F_1(\Sigma_3 + ...) \dots F_p(\Sigma_3 + ...) \big\rangle_Y \;.
	$$
(Although ${\rm Lie}(G)$ and ${\cal P}$ are vector spaces over
${\mathbb C}$, it is understood from Section \ref{sec:motiv} that the
bosonic degrees of freedom in $X$ and $Y$ are integrated only over the
{\it real} directions specified by the tangent spaces of ${\bf
X}_{A_1|A_1}$ and its projection ${\rm U}(1,1|2) / {\rm U}(1|1) \times
{\rm U}(1|1)$.)  Because $X$ does not appear in $Q$, the integral over
$X$ still gives ${\rm Det}(K)$, which is cancelled by the inverse
determinant from the bosonic ghost.  The remaining integral over $Y$
is more complicated, but will work out to give the ``right'' answer.
The reason is that the very same integral is obtained when evaluating
the BRST-invariant correlator (\ref{cf}) in Pruisken's nonlinear sigma
model.  In the latter case, too, the BRST mechanism reduces the
functional integral to small fluctuations around $Q(x) = \Sigma_3$.
In Gaussian approximation, the action functionals of Pruisken's model
and the $A_1|A_1$ model formally agree (on setting $Q = g \Sigma_3
g^{-1}$), and we can make them coincide by putting $\sigma_{xx} =
(8\pi f^2)^{-1}$.  Hence the dimensionally reduced theories are the
same, and we conclude
	$$
	\Big\langle F_1(Q(x_1))\dots F_p(Q(x_p))\Big\rangle_{A_1|A_1} =  
	\Big\langle F_1(Q(x_1))\dots F_p(Q(x_p))\Big\rangle_{\rm Pruisken}
	\;.
	$$
Thus the question whether BRST-invariant correlators of the form
(\ref{cf}) are trivial and are correctly normalized in the $A_1|A_1$
model, reduces to the same question in Pruisken's theory.  We may take
it for granted that the answer to the latter question is in the
affirmative.

In summary, requirements 3 and 4 are satisfied if the $A_1|A_1$ model
is augmented by a bosonic ghost in the manner described above.
Moreover, the mechanism at work suggests that physical observables for
the disordered electron gas are those which can be expressed in terms
of Pruisken's field $Q = g \Sigma_3 g^{-1}$.
	
\section{Conformal invariance}
\label{sec:conformal}

With the normalization issue taken care of, we now turn to requirement
2: conformal invariance.  This point has already been argued
convincingly in two recent papers for the target spaces ${\rm
PSU}(2|2)$ and ${\rm PSL}_{\mathbb R}(2|2)$, so we can afford to be
brief.  In Ref.~\cite{bvw} the action functional of the ${\rm
PSU}(n|n)$ nonlinear sigma model was related to that for ${\rm
U}(n|n)$.  It was then shown that the term
	$$
	S_0 = (2\pi f^2)^{-1} \int d^2 x \,
	\langle g^{-1}\partial g , g^{-1} \bar\partial g \rangle
	$$ 
does not renormalize in ${\rm U}(n|n)$, and hence not in ${\rm PSU}
(n|n)$, for any value of the coupling constant $f$.  The vanishing of 
the beta function for $S_0$ had already been pointed out by Gade and
Wegner in their 1991 paper \cite{gw} on ${\rm U}(m)$ at $m = 0$, which
is perturbatively equivalent to ${\rm U} (n|n)$.  The result can be
understood qualitatively from the fact that the beta function of $S_0$
for ${\rm U}(n|n)$ is independent of $n$, and for $n = 1$ the gauged
theory ${\rm PSU}(1|1)$ is free and finite.  In Ref.~\cite{bzv},
essentially the same nonlinear sigma model [with target space ${\rm
PSL}_{\mathbb R}(2|2)$] was considered.  There, conformal invariance 
was attributed to the quadratic Casimir invariant being zero in 
the adjoint representation.  The vanishing of the Casimir causes the 
vanishing of certain invariant current-current correlation functions,
which in turn protects the component $T_{z\bar z}$ of the
energy-momentum tensor (which is zero in the classical theory) from
becoming nonzero as a result of quantum fluctuations.  These arguments
are independent of the signature of the metric of the manifold and
apply equally to our case.  They are also robust enough to accommodate
the presence of a WZW term.  Thus the important message is that we
have a family of conformal field theories with {\it two} parameters
$f$ and $k$ to play with.  The Virasoro central charge of these
theories is easily seen \cite{bvw,bzv} to be $c = -2$.

By standard manipulations done on the Lagrangian, the holomorphic 
part of the energy-momentum tensor has the classical form
	$$
	T(z) = {\textstyle{1\over 2}} f^2 \big\langle g^{-1} 
	\partial g , g^{-1} \partial g \big\rangle + {\textstyle
	{1\over 2}} f^2 \, \partial\bar\varphi \, \partial\varphi \;, 
	$$
which is independent of the topological coupling $k$ and is not 
changed by quantum fluctuations.  On general grounds \cite{bpz}, the 
singular part of the operator product expansion of $T(z)$ with itself 
must be of the standard form
	$$
	T(z) T(w) = {c \over 2(z-w)^4} + {2 \over (z-w)^2} T(w)
	+ {1 \over z-w} \partial_w T(w) + ... \;.
	$$
The Virasoro central charge of a complex free boson (or of a pair of
real free bosons) is $+2$.  Together with the contribution from the
sigma model field $g$, this gives a total central charge of $c = -2 +
2 = 0$ in our case.  (The central charge vanishes here for the trivial
reason that the field theory partition function always equals unity,
independent of the size of the system.  A more informative quantity
can be introduced by following a recent proposal by Gurarie
\cite{gurarie}.)

In \cite{bzv} it was emphasized that the chiral symmetry algebra of
the theory is not just Virasoro but is actually much larger.  While
only the case $k = 0$ was discussed in that reference, the reasoning
easily carries over to the general case, as follows.  By varying the
action functional $S = S_0 / f^2 + ik \Gamma$, we obtain the equation
of motion $\partial_\mu J_\mu = 0$, where the conserved (left-invariant) 
current is
	\begin{equation}
	J_\mu = {1 \over 4\pi} \left( f^{-2} \, g^{-1} \partial_\mu g 
	+ ik \epsilon_{\mu\nu} g^{-1} \partial_\nu g \right) \;.
	\label{current}
	\end{equation}
To keep the following equations simple, we introduce a current
one-form ${\cal J}$ by
	$$
	{\cal J} = \alpha_1 \star g^{-1} {\rm d}g + i\alpha_2
	\, g^{-1} {\rm d}g \;,
	$$
where $\alpha_1 = 1/4\pi f^2$, $\alpha_2 = k / 4\pi$, ${\rm d}g =
\partial g \, {\rm d}z + \bar\partial g \, {\rm d} \bar z$, and the
star operator acts by $\star {\rm d}z = -i {\rm d}z$, $\star {\rm d}
\bar z = i {\rm d}\bar z$.  The equation of motion for $J_\mu$ then
translates into the statement that ${\cal J}$ is closed: ${\rm d}
{\cal J} = 0$.  (Note that ${\cal J}$ is the Hodge dual of what is
usually understood to be the current.)  Inverting the expression
for ${\cal J}$, we have
	$$
	g^{-1} {\rm d}g = (\alpha_1^2 - \alpha_2^2)^{-1} \left( - 
	\alpha_1 \star {\cal J} + i\alpha_2 {\cal J} \right) \;,
	$$
and differentiation (together with ${\rm d}{\cal J} = 0$) gives the 
integrability condition
	$$
	[ g^{-1} \bar\partial g , g^{-1} \partial g ] = 
	- {i\alpha_1 \over \alpha_1^2 - \alpha_2^2 } \left(
	\partial {\cal J}_{\bar z} + \bar\partial {\cal J}_z \right) \;,
	$$
for the components defined by ${\cal J} = {\cal J}_z {\rm d}z + {\cal
J}_{\bar z} {\rm d}\bar z$.  By combining this with the equation of
motion $\partial{\cal J}_{\bar z} = \bar\partial{\cal J}_z$, we arrive at
	$$
	\bar\partial {\cal J}_z = {\textstyle{i \over 2}} \left( \alpha_1
	- \alpha_2^2 / \alpha_1 \right) [ g^{-1} \bar\partial g ,
	g^{-1} \partial g ] \;.
	$$
We see that the current ${\cal J}_z$ is holomorphic for $\alpha_1 =
\alpha_2$ (or $1/f^2 = |k|$), as required for the current of a WZW
model.  Away from that limit, the current is not holomorphic.
However, the discrepancy is a {\it commutator}, so if $t_{a_1 a_2
... a_n}$ is any one of the large set of invariant symmetric tensors
for ${\rm psl}(2|2)$, the corresponding polynomial current
	$$
	W_{[t]} = t_{a_1 a_2 ... a_n} 
	{\cal J}_z^{a_1} {\cal J}_z^{a_2} ... {\cal J}_z^{a_n}
	$$
is holomorphic: $\bar\partial W_{[t]} = 0$, at least at the classical 
level.  It is expected \cite{bzv} that $W_{[t]}$ remains holomorphic 
in the quantum theory.

\section{More checks}
\label{sec:checks}

We now check the next three points on our list.  Of these, numbers 5
and 7 have already been input in the course of our development, so we
will be brief.  The former says that the fixed-point theory must have
a global chiral symmetry ${\rm SL}(2|2)_{\rm L} \times {\rm SL}(2|2)_
{\rm R}$, and the central subgroup of ${\rm SL}(2|2)$ generated by the
unit matrix has to be represented as a gauge degree of freedom. Both
features are automatic by the construction of the target space ${\bf
X}_{A_1|A_1}$ and the choice of Lagrangian made.  Note also that,
although ${\rm GL}(2|2)$ does not act on ${\rm SL}(2|2)$, it does act
on $Q = g\,\Sigma_3\,g^{-1}$ ($g \in {\rm SL}(2|2)$) by conjugation,
so the symmetries of Pruisken's nonlinear sigma model are fully
present.

Number 7 requires the density of states to be noncritical.  In
Pruisken's model, this observable is represented by the operator ${\rm
STr}\,\Sigma_3 Q$.  We have already identified the field $Q$ with
${\rm Ad} (g) \Sigma_3$ in the $A_1|A_1$ theory.  The field $Q$, and
the density of states along with it, will be noncritical if its
scaling dimension is zero.  The latter is in fact the case here.  On
general symmetry grounds the scaling dimension of an operator must be
expressed by the corresponding Casimir eigenvalues, and it can be
shown \cite{bzv} that all Casimir invariants vanish in the adjoint
representation of ${\rm psl}(2,2)$.

Requirement 6 concerns the fate of the theory under a perturbation
	$$
	{\rm STr} \, \omega \left( {\rm Ad}(g) - 1 \right) \Sigma_3
	$$ 
with $\omega = {\rm diag}(a_0 , 1 , b_0 , 1)$.  As can be seen, for
example from the result (\ref{dets}), such a perturbation gives a mass
$-\ln(a_0 b_0)$ to some fields in the Boson-Boson sector, and masses
$-\ln(a_0)$ or $-\ln(b_0)$ to some of the fermionic fields.  In the
infrared limit, these degrees of freedom disappear from the theory.
What is left behind is the Fermion-Fermion sector, which remains
massless, being protected by a residual ${\rm SU}(2)_{\rm L} \times
{\rm SU(2)} _{\rm R}$ symmetry.  Restriction of the theory to this sector
yields a Lagrangian with WZW term and topological coupling $k$.
Requirement 6 then says that the integer $k$ has to take its smallest
nonzero value:
	\begin{equation}
	k = 1 \;.
	\label{level}
	\end{equation}
This determines one of the two coupling constants of the theory.
The correct choice of the other coupling $f$ is less obvious.

\section{The marginal coupling $f$}
\label{sec:coupling}

Renormalization group fixed points in conventional chiral nonlinear
sigma models with WZW term, come as a discrete one-parameter family:
once a value for the level $k$ has been chosen, the other coupling has
to be $f^2 = 1/|k|$ in order for the theory to be conformally
invariant.  In the present case the situation is different.  Given a
value for $k$, conformal invariance does not determine $f$, which is a
truly marginal coupling parametrizing a {\it line} of fixed points
$f^2 \le 1/|k|$.

While being an intriguing feature, the existence of a truly marginal
perturbation is a threat to universality and therefore problematic for
our proposal.  Unless there exists some hidden symmetry or other
constraint, it stands to reason that the marginal direction can be
explored by the disordered electron gas: since the renormalization
group trajectories of different members of the QH family originate
from different initial conditions, we expect the critical trajectories
to terminate on different points on the fixed line, leading to a
variety of critical behavior.  In contradistinction, the critical
properties seen in numerical and real experiments on the QH transition
appear to have a high degree of universality.  The latter seems to
suggest that the critical behavior is governed by a single fixed
point, not a one-parameter family of such points.

The question, then, is how the observed universality can be reconciled
with the marginality of $f$.  What we need is some mechanism to ensure
that the critical RG trajectories for different members of the QH
universality class intersect the fixed line at the same point.  The
scenario we offer is this.  The coupling constant $f$ determines the
most singular terms (proportional to unity) of the operator product
expansion for the vertex operators of the field theory.  Thus the
value of $f$ controls the short-distance singularities.  In
particular, $f$ controls the fluctuations of the field $g$ at short
distances from a conducting (or absorbing) boundary.  At the same
time, the short-distance physics of the disordered electron gas near a
conducting boundary is {\it classical diffusion} of electrons.  As is
well understood, the conductance and other observables can be
expressed as sums over electron paths.  In a typical path, loops are
prevalent in the bulk of the system, but rare in the vicinity of an
absorbing boundary.  Thus, while the dynamics in the bulk is strongly
influenced by quantum interference effects (due to loops) and
incipient localization, interference near a boundary is cut off by
absorption (or exiting probability flux) and the motion can be treated
as being classical there.  (``Classical'' here means that the absolute
square of a coherent sum of path amplitudes is well approximated by an
incoherent sum of squares.)  In the field theory, classical diffusion
corresponds to small field fluctuations around a ``vacuum'' selected
by the absorbing boundary conditions.  This correspondence is, in
fact, what leads to the identification, in Pruisken's theory, of the
coupling $\sigma_{xx}$ with the classical longitudinal conductivity
(which in turn is proportional to the classical diffusion constant by
the Einstein relation).  The same identification will be made for
$1/8\pi f^2$ in the $A_1|A_1$ model.  By this argument, the parameter
$f$ is completely determined by matching to the classical dynamics
near a conducting boundary.  Universality is therefore guaranteed to
the extent that the classical diffusion constant has a universal
value.  Evidence in favor of the latter is provided by the semicircle
relation
	\begin{equation} 
	\sigma_{xx}^2 + (\sigma_{xy} - 1/2)^2 = 1/4\;, \label{semic} 
	\end{equation} 
which has been argued to be universally valid for incoherent QH
systems \cite{rf}.  For recent developments related to this subject
see \cite{stsss,sss,shewe}.

Our strategy in the sequel will be as follows.  As was first pointed
out in \cite{iqhe}, charge transport in a critical conductor is probed
least obtrusively by measuring a conductance between {\it interior}
contacts.  The simplest theoretical setup is to take an infinite
system with two interior contacts that are small.  At criticality, the
conductance will then be some algebraically decaying two-point
function of conformal field theory.  The critical exponent determining
the decay at large distances between the two contacts is expected to
be universal, but nontrivial to compute.  (If the contact in
field-theoretic representation is expanded in scaling fields with
scaling dimension $\Delta_\lambda$, the exponent equals $2 {\rm Min}
\Delta_\lambda$.)  On the other hand, the form of the short-distance
singularity of the conductance depends on the size or ``strength'' of
the contacts.  If the contacts are strong, so that an electron near a
contact quickly exits and the time spent in the system is short, the
conductance becomes {\it classical} at small distances between the two
contacts.  (In contrast, for contacts that are point-like or weak, the
classical transport regime does not exist, as the behavior of the
conductance is controlled by quantum interference even at the smallest
distances.)  What we will do is to extract from the field theory the
short-distance singularity of the conductance between two strong
contacts.  Matching the result to the classically expected singularity
will then determine $f$.

\subsection{Classical point-contact conductance}
\label{sec:classical}

The first thing to describe is the classical expectation.  We will
work in the infinite plane ${\mathbb R}^2 \simeq {\mathbb C}$ with
complex coordinate $z = x_1 + ix_2$.  From linear response theory, the
conductance, $G$, is a current-current correlation function,
	\begin{equation}
        G = \oint\limits_{z\in C_1} \oint\limits_{z^\prime\in C_2}
	\big\langle j(z) j(z') \big\rangle \;, \label{condg}
	\end{equation}
where, for a system with interior contacts at points with coordinates 
$z_1$ and $z_2$, the integration contours $C_1$ and $C_2$ are two 
disjoint cycles enclosing the points $z_1$ resp.~$z_2$ (Figure 2).  
\begin{figure}
	\hspace{2cm} 
	\epsfxsize=10cm \epsfbox{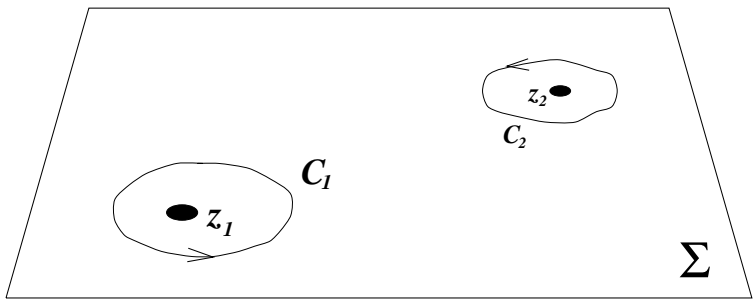}
        \caption{Conductance is a current-current correlation function.}
        \label{fig:pcontac}
      \end{figure}
For a {\it classical} conductor, the current one-form $j$ can be
expressed in terms of a real boson field $\varphi$ by
	$$
	j = {1\over i} \left( {\partial\varphi\over\partial z} {\rm d}z 
	- {\partial\varphi\over\partial\bar z} {\rm d}\bar z \right) 
	= \epsilon_{\mu\nu} \partial_\mu \varphi \, {\rm d}x_\nu \;,
	$$
and the expectation value $\langle \bullet \rangle$ is defined by a
Gaussian functional integral
	$$
	\big\langle\bullet\big\rangle = {1 \over Z} \int {\cal D}\varphi 
	\;\bullet\; \exp \int d^2x \; \varphi\partial\bar\partial\varphi \;.
	$$
(We are working in scaled units here and will put back the proper
units later.)  As usual, the partition function $Z$ normalizes the 
integral, so that $\langle 1 \rangle = 1$.  The escape of probability
flux through the contacts is modeled by imposing ``absorbing''
boundary conditions at these points:
	$$
	\varphi(z_1) = \varphi(z_2) = 0 \;.
	$$
By the equation of motion for $\varphi$ ($\bar\partial\partial
\varphi = 0$), the current is conserved: ${\rm d} j = 0$ on 
${\mathbb C} \setminus \{ {z_1} \cup {z_2} \}$, and the expression
(\ref{condg}) does not change under small deformations of $C_1$ and
$C_2$.

We now wish to know how the conductance $G$ depends on $|z_1-z_2|$,
the distance between the two contacts.  One way of computing the
dependence is to use an analogy with classical 2d electrostatics, by
reinterpreting $\varphi$ as an electric potential, $j$ as an electric
flux {\it etc.}  We can then view the functional integral average
$\langle \varphi(z) \varphi (z^\prime) \rangle$ as the Green
function of a Poisson problem, which can be solved by the technique of
conformal mapping.  Alternatively we can compute $G$ directly, by
manipulation of the functional integral in the way detailed in the
next subsection.

However, we will take the short cut of guessing the answer on
dimensional grounds.  Because the components of a conserved current
have scaling dimension one, and a coordinate differential subtracts
one dimension, the scaling dimension of the current one-forms $j(z)$
integrated in (\ref{condg}) is zero.  By conformal invariance of the
free boson theory, the conductance $G$ can therefore depend on $|z_1 -
z_2|$ only through the logarithm.  On physical grounds, $G$ must {\it
decrease} with increasing distance between the contacts, so we expect
$G \sim (\ln |z_1 - z_2|)^{-1}$.  The constant of proportionality can
be determined by comparison with the quasi-1d limit.  According to
Ohm's law, the conductance of a cylinder of circumference $W$, height
$|z_1 - z_2|$, and conductivity $\sigma$, is $G = \sigma W / |z_1 -
z_2|$.  On replacing the (quasi-)1d Coulomb propagator $|z_1 - z_2| /
2W$ by the 2d Coulomb propagator $(2\pi)^{-1} \ln |z_1 - z_2|$, we
arrive at
	\begin{equation}
	G_{2{\rm d}}^{\vphantom{\dagger}} = 
	{\pi\sigma \over \ln (|z_1-z_2|/R)} \;,
	\label{sdist}
	\end{equation}
which is the correct result.  A length scale $R$ was inserted in the
argument of the logarithm to get the physical dimensions straight.
(We may regard $R$ as the size of the contacts, which sets the scale
and serves as a short-distance cutoff.)

\subsection{Classical conductance from field theory}
\label{sec:condWZW}

Our next task is to reproduce the classical result (\ref{sdist}) from
the short-distance physics of the $A_1|A_1$ model.  As before, we
consider an infinite system with interior point contacts placed at two
positions $z_1$ and $z_2$, and compute the conductance as a
current-current correlation function.

To begin, recall from the end of Section \ref{sec:brst} the
parametrization $g(x) = {\rm e}^{Y(x)} {\rm e}^{X(x)} g_0$, where
$X(x)$ takes values in ${\rm Lie}(H)$, and $Y(x)$ in the complement
${\cal P}$ of ${\rm Lie}(H)$ in ${\rm psl}(2|2)$.  The latter
determines the field of Pruisken's model by $Q = {\rm e}^Y \Sigma_3
{\rm e}^{-Y} = {\rm e}^{2Y} \Sigma_3$, while the former is
post-Pruisken.  From the supersymmetric nonlinear sigma model for
disordered metals \cite{efetov}, the boundary condition for $Q$ is
known to be $Q = \Sigma_3$ on any well-conducting boundary. By
transcription to the present case, we require
	$$
	Y(z_1) = Y(z_2) = 0 \;.
	$$
The linear field $X$ is {\it not} subject to any such boundary
condition.  However, in the limit of vanishing regularization
parameter $\epsilon \to 0$, we can exploit the global invariance of
the field theory to set $X = 0$ at {\it one} of the two contacts, say
$z_1$.  At that contact, we then have
	$$
	g(z_1) = g_0 \;.
	$$
(We here ignore the subtle issue whether such a boundary condition is
admissible in the presence of a Wess-Zumino term.)

By the use of current conservation, we may shrink the contours $C_1$
and $C_2$ to infinitesimal loops encircling the contacts $z_1$ and
$z_2$.  In the vicinity of the first contact $z_1$, the field $g$
performs small fluctuations around the identity coset $g_0$.
Therefore, if the second contact $z_2$ is close enough to the first
one, we may compute the current-current correlation function by
expanding $g$ in $X$ and $Y$, and truncating at quadratic order.  (Of
course, the theory becomes nonlinear far from $z_1$, but the
contributions from there cancel by supersymmetry and $c = 0$.)  Thus
the problem reduces to a free-field calculation.

From the variation of the action functional (\ref{action}), we obtain
the expression (\ref{current}) for the conserved current of the
$A_1|A_1$ model.  That current is ${\rm psl}(2|2)$-valued.  Linear
response theory, in its transcription to the field-theoretic formalism
for disordered electron systems, instructs us to expand the current in
${\rm psl} (2,2)$ generators: $J_\mu = {\textstyle{1 \over 2}} \sum_a
J_\mu^a T^a$, and pick any component $J_\mu^a$ whose generator $T^a$
{\it anticommutes} with the matrix $\Sigma_3$.  Since the diagonal
matrix $\Sigma_3$ distinguishes between the advanced and retarded
sector, such a component of the conserved current mixes retarded and
advanced degrees of freedom, and its correlation function gives the
conductance.  To be definite, we take $J_\mu^a$ to be a bosonic
component of $J_\mu$ , and the corresponding generator has the
standard normalization $\langle T^a , T^a \rangle = 2$.

Because it suffices to do a free-field calculation, we may 
approximate
	$$
	J_\mu^a \approx {1\over 8\pi} \left( f^{-2} \partial_\mu
	Y^a + ik \epsilon_{\mu\nu} \partial_\nu Y^a \right) \;,
	$$
where $Y^a$ is determined by $Y = {\textstyle{1\over 2}} \sum_a Y^a
T^a$.  On integration along one of the cycles $C_i$, the first term
yields the radial current flow emanating from or sinking into a
contact, whereas the second term is topological and measures the
vorticity of the flow around a contact.  By the choice of boundary
condition $Y(z_1) = Y(z_2) = 0$, the topological current is exact, and
the vorticity must therefore vanish, so we may neglect the latter
term. (Note that the topological current does have a finite effect in
the presence of an {\it insulating} boundary \cite{xrs}.)  Hence, if
we simplify the notation by putting $\varphi \equiv Y^a$, we have
$J_\mu^a = (8\pi f^2)^{-1} \partial_\mu\varphi$. To obtain the total
current through a contour $C_i$, we need to integrate the (Hodge dual
of the) current one-form,
	$$
	j \equiv \epsilon_{\mu\nu} J_\mu^a {\rm d}x_\nu = 
	(8\pi f^2)^{-1} \epsilon_{\mu\nu} \partial_\mu\varphi 
	\, {\rm d}x_\nu \;.
	$$
With this identification, the conductance is again given by
(\ref{condg}), the functional integral now being
	$$
	\big\langle\bullet\big\rangle = {1 \over Z} \int {\cal D}\varphi 
	\;\bullet\; \exp {1 \over 4\pi f^2} \int d^2x \; 
	\varphi\partial\bar\partial\varphi \;.
	$$
The normalizing denominator arises from integrating over the
components of $Y$ other than $Y^a$ (while integration over $X$ cancels
the determinant from the bosonic ghost).  Hence we are back to the
problem considered in the previous subsection.  Instead of guessing
the answer for $G$, let us now compute the conductance explicitly,
carrying along the necessary scale factors.

The first step is to make the boundary conditions explicit in the
functional integral.  For that we place Dirac $\delta$--distributions
on the field $\varphi$ at $z_1$ and $z_2$.  The ``partition function''
$Z$ then takes the form
	$$
	Z = \int {\cal D}\varphi \; \delta[\varphi(z_1)] \;
	\delta[\varphi(z_2)] \; \exp {1 \over 4\pi f^2} \int d^2x \; 
	\varphi \partial\bar\partial\varphi \;,
	$$
where $\delta [\varphi] = \lim_{\varepsilon\rightarrow 0} (\pi
\varepsilon)^{-1/2} {\rm e}^{-\varphi^2/\varepsilon}$. A similar 
formula holds for the numerator of the current-current correlation
function.

The next step is to ``fuse'' the $\delta$--distribution $\delta
[\varphi(z_1)]$ with the operator $O_1[\varphi] \equiv \oint_{z \in
C_1} j(z)$, by contracting $C_1$ to the point $z_1$ (and similarly
with $C_2$ at $z_2$) to produce a {\it local} operator.  The dominant
term of the operator product expansion is extracted as follows. We put
$z - z_1 = {\rm e}^{\tau +i\sigma}$ ($\tau , \sigma$ are the
coordinates that are used for what is called ``radial quantization''
around the point $z_1$), and take $C_1$ to be a circle $C_1 = \{ z \in
{\mathbb C} : |z-z_1| = \varepsilon \}$.  Then
	$$
	O_1[\varphi]={1\over 8\pi f^2}\oint\limits_{\tau=\ln{\varepsilon}}
	{\partial\varphi\over\partial\tau} d\sigma\;.
	$$
To go further we use a simple analogy. Consider a quantum particle in
one dimension with mass $m$ and coordinate $x$. Then we know that the
velocity $\dot x = dx / d\tau$ in Feynman's imaginary-time path
integral translates into the operator $m^{-1} d/dx$ in Schr\"odinger
quantum mechanics.  Moreover, for a free particle,
	$$
	\int\limits_{x(0) = x_i}^{x(T) = x_f} {\cal D}x \;
	{\rm e}^{-(m/2)\int\limits_0^T \dot x^2 d\tau} \dot x(\tau_0) =
	\left< x_f \left| \exp\left( \frac{T}{2m}\frac{d^2}{dx^2}
	\right) m^{-1} d/dx \right| x_i\right>\;,
	$$
independent of the time $\tau_0$ where the velocity is evaluated.
Similarly, if $\varphi_i(\sigma)$ and $\varphi_f(\sigma)$ are
prescribed functions ${\rm S}^1\rightarrow {\mathbb R}$, then 
in the present problem we have
	\begin{eqnarray*}
	&&\int\limits_{\varphi(\bullet,T_i) = \varphi_i}^
	{\varphi(\bullet,T_f) = \varphi_f} {\cal D}\varphi 
	\left( \oint_{\tau = \ln{\varepsilon}}
	\frac{\partial\varphi}{\partial\tau} d\sigma \right)
        \exp {1 \over 4\pi f^2} \int\limits_A d^2x \; 
	\varphi\partial\bar\partial\varphi \\ 
	&=& 8\pi f^2 \oint d\sigma \left\langle \varphi_f \left| 
	{\rm e}^{-(T_f-T_i){\cal H}} \frac{\delta}{\delta\varphi
	(\sigma)}\right| \varphi_i\right\rangle \;.
	\end{eqnarray*}
Here $A$ denotes the annulus $T_i< \tau < T_f$, $0 \le \sigma < 2\pi$,
and ${\cal H}$ is the Hamiltonian of the radially quantized theory. Thus,
by shrinking $C_1$ to the point $z_1$ we obtain
	$$
	O_1[\varphi] \; \delta[\varphi(z_1)] \; \longrightarrow \; 
	\frac{\partial}{\partial\varphi(z_1)} \; \delta[\varphi(z_1)]
	\equiv \delta^\prime [\varphi(z_1)] \;.
	$$
Doing the same at the other contact and returning from the radially
quantized theory to the functional integral, we arrive at the expression
	$$
	G = \frac{1}{Z}\int {\cal D}\varphi \;
	\delta^\prime [\varphi(z_1)] \; \delta^\prime [\varphi(z_2)] \;
	\exp {1\over 4\pi f^2}\int d^2x \;\varphi\partial\bar\partial\varphi\;.
	$$
This is easy to compute by Fourier expanding the $\delta$--distributions
in terms of ``vertex operators'':
	$$
	\delta[\varphi] = \int\limits_{\mathbb R} 
	{d\lambda\over 2\pi} \; \exp i \lambda \varphi \;.
	$$
In the case of the partition function $Z$, a standard result for Gaussian 
functional integrals gives
	\begin{eqnarray*}
	Z &=& \int\limits_{\mathbb R}{d\lambda \over 2\pi}
	\int\limits_{\mathbb R} {d\lambda^\prime \over 2\pi} 
	\int{\cal D}\varphi \; \exp \left( (4\pi f^2)^{-1}\int d^2x 
	\; \varphi\partial\bar\partial\varphi + i\lambda\varphi(z_1) 
	+ i\lambda^\prime\varphi(z_2) \right) \\
	&=& {\rm const} \times \int_{\mathbb R} {d\lambda \over 2\pi} 
	\; \exp - 2\pi f^2 \lambda^2 \, K(z_1,z_2) \;,
	\end{eqnarray*}
where $K$ is the Coulomb propagator, $K(z_1,z_2) = (\partial\bar\partial)
^{-1}(z_1,z_2)$. Doing the same computation for the numerator of the
current-current correlation function and taking the ratio, we find
	$$
	G = \left( 4\pi f^2 K(z_1,z_2) \right)^{-1} = 
	{1 / 8f^2 \over \ln(|z_1-z_2|/R)} \;.
	$$
From the previous subsection, we know that the numerator of the last
expression equals $\pi \sigma_{xx}$, with $\sigma_{xx}$ the classical
conductivity.  Hence, we have the result
	\begin{equation}
	f^2 = (8\pi\sigma_{xx})^{-1} \;.
	\label{coupling}
	\end{equation}
If we appeal to the semicircle relation (\ref{semic}), which gives
$\sigma_{xx} = 1/2$ for the critical value $\sigma_{xy} = 1/2$, we get
$f^2 = 1/4\pi$.  Note that for this value of the coupling the
current (\ref{current}) is not holomorphic, and the field $g$ does
not separate into left-moving and right-moving waves.  This will make 
it harder to obtain analytical solutions than in the conventional
(affine Lie algebra) case.

\section{Quantum point-contact conductance}
\label{sec:quantum}

With both couplings of the $A_1|A_1$ model now determined, we put the
theory to test on a nontrivial transport coefficient: the conductance
between two interior contacts in the {\it quantum limit} of point-size
contacts.  This observable was the object of a recent study \cite{jmz}
where heavy analytical machinery was combined with numerical
simulation.  Starting from the Chalker-Coddington model with two point
contacts separated by a distance $r$, the following expression for the
$q$-th moment of the point-contact conductance was derived:
	\begin{equation}
	\big\langle G^q \big\rangle = \int_0^\infty d\mu(\lambda)
	\, C_q(\lambda) \, r^{-2\Delta_\lambda} \;, \label{pcontac}
	\end{equation}
where
	$$
	d\mu(\lambda) = {\lambda\over 2} \tanh \left( {\pi\lambda
	\over 2} \right) d\lambda \;, \qquad
	C_q(\lambda) = \left| { \Gamma\left(q-{1+i\lambda\over 2}\right) 
	\over \Gamma(q)} \right|^2 \;.
	$$
(Note that the conductance of a phase-coherent quantum system {\it
fluctuates} as a function of disorder \cite{janssen,ck}, and for a
complete description we need to compute the entire distribution
function or, equivalently, all of the moments of the conductance.
Note also that the use of the symbol $\lambda$ is not accidental: this
parameter plays the same role here as in the classical calculation of
subsection \ref{sec:classical}.) The factors $d\mu(\lambda)$ and
$C_q(\lambda)$ are purely kinematical, the first one being a
Plancherel measure and the second the square of a Clebsch-Gordan
coefficient, both of which are determined by representation theory and
harmonic analysis.  All dynamical information from the field theory
resides in the $r$-dependent factor $r^{-2 \Delta_ \lambda}$.  The
exponents $\Delta_\lambda$ are the scaling dimensions of certain
scaling fields $\phi_\lambda$.  To elucidate their origin, we now give
a rough summary of how the formula for $\langle G^q \rangle$ is
obtained (the details are found in \cite{jmz}).

The first step is to cast the Chalker-Coddington network model in the
form of a supersymmetric vertex model \cite{network} with global ${\rm
GL}(2|2)$ symmetry, by using the color-flavor transformation in the
way indicated in Section \ref{sec:cc}.  The partition function of the
vertex model is a sum over (classical) degrees of freedom situated on
the links of the network.  They take values in the ${\rm GL}(2|2)$
modules $V$ and $V^*$ that were defined in Section \ref{sec:spin}.  In
the vertex-model representation, the point-contact conductance assumes
the form of a two-point correlation function,
	$$
	\big\langle G^q \big\rangle = \big\langle O_q^+(0) \,
	O_q^-(r) \big\rangle \;,
	$$
where $O_q^\pm$ are operators that are viewed as elements of $V \otimes 
V^*$, and are given in terms of the vacuum $|0\rangle$ and the superspin 
generators $S_{ij}$ of Section \ref{sec:spin} by
	$$
	O_q^+ = (S_{02})^q | 0 \rangle \otimes | \bar 0 \rangle \;,
	\qquad
	O_q^- = | 0 \rangle \otimes (S_{20})^q | \bar 0 \rangle \;.
	$$
Precisely speaking, to make sense of the expression for $\langle G^q
\rangle$, one canonically identifies $V\otimes V^*$ with ${\rm End}
(V)$ (using that $V^*$ is dual to $V$, see Section \ref{sec:spin}),
and computes the vertex model sum with $O_q^\pm \in {\rm End}(V)$
acting on the superspin at the corresponding link.

The second step is to Fourier-analyze the elements $O_q^\pm$ of the
tensor product $V \otimes V^*$.  Because the vertex model has global
${\rm SL}(2|2)$ symmetry, one wants to decompose $V \otimes V^*$ into
irreducibles w.r.t.~this group or, with no loss, w.r.t.~${\rm PSL}
(2|2)$.  This decomposition was first described in Section 5.2 of
\cite{iqhe}.  It turns out that the only representations that appear
are those of a {\it continuous} series, labelled by $\lambda \in
{\mathbb R}^+$, which is essentially the same as the principal
continuous series of unitary representations of ${\rm SU}(1,1)$ --- 
the textbook example for harmonic analysis on a noncompact group.  
In a self-explanatory notation, the Fourier decomposition of the 
operators $O_q^\pm$ is written
	\begin{equation}
	O_q^+ = \int_0^\infty d\mu(\lambda) \, \big\langle Vq , V^* 0 
	| \lambda q \big\rangle \, \phi_{\lambda q} \;,
	\qquad O_q^- = \overline{O_q^+} \;. \label{fourier}
	\end{equation}
These decompositions are inserted into the correlation function $\big
\langle O_q^+(0) \, O_q^-(r) \big\rangle$.  Adopting the phase
convention $\overline{\phi_{\lambda q}} = \phi_{\lambda -q}$, denoting 
the product of Clebsch-Gordan coefficients by
	$$
	C_q(\lambda) = \Big| \big\langle Vq , V^* 0 | 
	\lambda q \big\rangle \Big|^2 \;,
	$$
and postulating
	\begin{equation}
	\big\langle \phi_{\lambda q} (0) \, \phi_{\lambda^\prime -q}(r)
	\big\rangle = {\delta(\lambda-\lambda^\prime) \over m(\lambda)}
	\, r^{-2\Delta_\lambda} \;, \label{2point}
	\end{equation}
where $m(\lambda)$ is defined by $d\mu(\lambda) = m(\lambda) d\lambda$, 
we then arrive at (\ref{pcontac}).

Let us pause here for a moment to insert the following remark. In 
\cite{bzv} the representation theory of ${\rm PSL}(2|2)$ was
discussed in some detail, with the focus being on {\it highest-weight}
representations.  The existence of representations without highest
weight was mentioned, but only parenthetically, as the authors of
\cite{bzv} ``... do not have the need for those''.  In our case, the
situation is quite the opposite!  Highest-weight representations (or,
for that matter, lowest-weight representations) play no role for the
conductance, and all the representations labelled by $\lambda$ are of
another type.  This feature is forced by the fact that the
representations of ${\rm PSL}(2|2)$ on the modules $V$ and $V^*$ are
{\it unitary} w.r.t.~to a subgroup ${\rm SU}(1,1) \times {\rm SU}(2)$,
so the decomposition of $V \otimes V^*$ into ${\rm PSL}(2|2)$
irreducibles is exhausted by unitary representations of this subgroup.
It is well known that the unitary representations of the noncompact
group ${\rm SU}(1,1)$ organize into discrete and continuous series.
All representations that occur in the tensor product $V \otimes V^*$
contain the vector $|0\rangle\otimes |\bar 0 \rangle \in V \otimes
V^*$, which is stable under the action of the compact ${\rm U}(1)$
subgroup of ${\rm SU}(1,1)$.  Thus, as far as ${\rm SU}(1,1)$ is
concerned, our problem amounts to doing harmonic analysis on ${\rm
SU}(1,1) / {\rm U}(1) \simeq {\rm H}^2$.  This eliminates the discrete
unitary series of ${\rm SU}(1,1)$ representations (all of which are
given by holomorphic sections of associated line bundles ${\rm
SU}(1,1) \times_{{\rm U}(1)} {\mathbb C}_{\mu}$ with {\it nonzero}
${\rm U}(1)$ charge $\mu$), leaving only the continuous series.  The
members of the latter are non-algebraic and do not contain any
highest-weight or lowest-weight vector.

Next, we indicate how to construct the functions $\phi_{\lambda q}$
appearing in the Fourier decomposition of $O_q^+ \in V \otimes V^*$.
For that we adopt the notation $G = {\rm PSU}(1,1|2)$ and $K = {\rm
PS}\left( {\rm U}(1|1) \times {\rm U}(1|1) \right)$.  By a
supersymmetric version of the Borel-Weil correspondence, the module
$V$ can be viewed as the space of holomorphic sections of a line
bundle $G \times_K {\mathbb C}_m$, while $V^*$ is the space of
antiholomorphic sections of the conjugate bundle $G \times_K {\mathbb
C}_{m^*}$.  (In fact, this is precisely how $V$ and $V^*$ first arose
in \cite{iqhe}.  $m$ is a one-dimensional representation of $K$ by the
superdeterminant, and $m^* = m^{-1}$.)  From this viewpoint, the
tensor product $V \otimes V^*$ lies inside some space of square
integrable functions on $G/K$, and the Fourier decomposition of $V
\otimes V^*$ becomes a sub-problem of the problem of harmonic analysis
on $G/K$.  (This is familiar from quantum mechanics: the modules $V$
and $V^*$ are Hilbert spaces that arise by geometric quantization of
the classical phase space $G/K$, and by multiplying the ``wave
functions'' of $V$ with the conjugate wave functions of $V^*$, we
recover $G/K$ in the form of ``Wigner functions'' with a fuzzy
resolution.  Note also the following.  To decompose the tensor product
$V \otimes V^*$, we need to diagonalize the low-order Casimir
invariants of $G$.  In the present case, consideration of the
quadratic Casimir turns out to be sufficient.  Under the embedding of
$V \otimes V^*$ into a function space on $G/K$, the quadratic Casimir
corresponds to a second-order differential operator invariant
w.r.t.~$G$, the Laplace-Beltrami operator.)  The latter problem was
solved by a supersymmetric adaptation of Harish-Chandra theory in
\cite{cmp,prl92,mmz}.  We now give a quick taste of the basic idea.

Let $G_{\mathbb C} = N_{\mathbb C} A_{\mathbb C} K_{\mathbb C}$ be an
Iwasawa decomposition \cite{helgason2} of $G_{\mathbb C} \equiv {\rm
PSL}(2|2)$.  (Compact groups such as ${\rm SU}(2) \subset G$ do not
admit an Iwasawa decomposition, so we are forced to work on the
complexification $G_{\mathbb C}$.)  The ``radial'' factor $A_{\mathbb
C}$ is a maximal abelian subgroup with Lie algebra contained in ${\cal
P}$ (where ${\cal P}$ is defined by ${\rm Lie}(G_{\mathbb C}) = {\rm 
Lie}(K_{\mathbb C}) + {\cal P}$).  The Iwasawa decomposition determines 
a radial function $A : G_{\mathbb C} \to {\rm Lie}(A_{\mathbb C})$ by
	$$
	g = n(g) \; {\rm e}^{A(g)} k(g) \;.
	$$
This function is left-invariant under $N_{\mathbb C}$, right-invariant
under $K_{\mathbb C}$, and restricts to a (complex-valued) function on
${\bf X}_{A_1|A_1}$.  Because $A_{\mathbb C}$ normalizes $N_{\mathbb
C}$, the radial part $L_A$ of the Laplace-Beltrami operator $L$ on $G$
is invariant under left translations by elements of $A_{\mathbb C}$.
As a result, $L_A$ is a differential operator with {\it constant
coefficients}, and its eigenfunctions are simple exponentials.  Thus,
any weight $\mu : {\rm Lie}(A_{\mathbb C}) \to {\mathbb C}$ gives rise
to an eigenfunction $\Phi_\mu(g) = \exp \mu(A(g))$ of $L_A$, and hence
of $L$.  (In general, such a function will be well-defined only
locally, and global consistency in the compact sector imposes an
integrality condition on $\mu$.)  With $\Phi_\mu(g)$, every rotated
function $\Phi_\mu(kg)$ $(k \in G)$ also is an eigenfunction of $L$.
The question addressed by harmonic analysis is which of the $\mu$ to
use in constructing the Fourier transform and its inverse, and what is 
the measure in Fourier space (the Plancherel measure). 

For the case at hand, the answer is known from \cite{mmz}. Since ${\rm
dim}(A_{\mathbb C}) = 2$, the weights are parametrized by two quantum
numbers: a continuous number $\lambda \in {\mathbb R}^+$ reflecting
the ${\rm SU}(1,1)$ content, and a discrete number $l \in 2{\mathbb
N}-1$ labelling representations of ${\rm SU}(2)$ with integer spin.
On restriction from the complete function space on $G / K$ to the
subspace $V \otimes V^*$, the second quantum number gets frozen at the
minimal value $l = 1$ (this apparently is related to the topological
coupling of the $A_1|A_1$ model being minimal), and we are left with a
single parameter $\lambda \in {\mathbb R}^+$.  The corresponding
eigenfunctions $\Phi_\lambda$ of the Laplace-Beltrami operator $L$ can
be constructed quite explicitly.  In the parametrization of $G/K$ by
the complex supermatrices $Z , \tilde Z$ introduced in (\ref{ztz}),
they are
	$$
	\Phi_\lambda = {\rm SDet} \left( {(1-\tilde Z)(1-Z) \over
	(1 - \tilde ZZ)} \right) ^{(1+i\lambda)/2} \;.
	$$
These are eigenfunctions of $L$ with eigenvalue $-(\lambda^2 + 1)$.
Other eigenfunctions with the same eigenvalue can be constructed by
acting with the symmetry group $G$ on $\Phi_\lambda$.  The set of
functions so obtained form an ``eigenfunction'' representation space,
$V_\lambda$, of $G$.  We will not go into the details of its
construction here.  Suffice it to say that the function $\phi_{\lambda
q}$ appearing in the Fourier decomposition (\ref{fourier}) is the
element of $V_\lambda$ with the same weights w.r.t.~the diagonal
generators $S_{ii} \otimes 1 + 1 \otimes S_{ ii}$ of $G$ as $O_q^+ \in
V \otimes V^*$.

The $\Phi_{\lambda q}$ extend to functions on ${\bf X}_{A_1|A_1}$ in
the natural way.  We can therefore consider the two-point correlator
$\langle \phi_{\lambda q}(0) \; \phi_{\lambda^\prime -q}(r) \rangle$
in the $A_1|A_1$ model.  On general grounds, such a correlation
function is of the form (\ref{2point}) for a conformal field theory.
Taking that formula for granted, the remaining problem is to compute
the scaling dimension $\Delta_\lambda$ of $\phi_{\lambda q}$.  By
$G$-invariance of the field theory, $\Delta_\lambda$ is the same for
every vector of the representation space $V_\lambda$.  In particular,
$\Delta_\lambda$ agrees with the scaling dimension of the function
$\Phi_\lambda$ generating the eigenfunction representation.  Leaving
aside the possibility of an exact determination, we can compute
$\Delta_\lambda$ from the algebraic decay of the correlator $\langle
\Phi_\lambda(0) \overline{\Phi_ {\lambda^\prime}}(r)\rangle$.  Since
$\Phi_\lambda$ is simply an exponential in Iwasawa coordinates,
perturbation theory is easy to implement and gives
	\begin{equation}
	\Delta_\lambda = f^2 (\lambda^2 + 1) \label{dim}
	\end{equation}
to leading order.  In Ref.~\cite{bzv} this result (for highest-weight
representations, but it doesn't make any difference) was conjectured
to be {\it exact}.  I have no nonperturbative proof of that conjecture
but also no evidence against it.  (As we are about to see, the
numerics of \cite{jmz} supports the conjecture.)  We shall therefore
{\it assume} (\ref{dim}) to be true, leaving its verification as an
open problem for future work.

Given the formulas (\ref{pcontac}) and (\ref{dim}), we have a complete
description of the quantum point-contact conductance.  For the purpose
of numerical computer simulation, the best observable to consider is
the {\it typical} conductance, $\exp \langle \ln G \rangle$.  This
statistic is determined by typical events (unlike the average
point-contact conductance, which is dominated by rare events), and can
therefore be computed with high statistical accuracy.  By
analytically continuing (\ref{pcontac}) from the positive integers $q$
to the vicinity of $q = 0$, and using $\langle G^q \rangle = 1 + q
\langle \ln G \rangle + {\cal O}(q^2)$, one obtains
	$$
	\big\langle \ln G \big\rangle = -2i {d \over d\lambda}
	r^{-2\Delta_\lambda} \big|_{\lambda = i} = - 8 f^2 \ln r
	$$
or upon exponentiation,
	$$
	\exp \big\langle \ln G \big\rangle = r^{-X_{\rm t}} \;,
	\qquad X_{\rm t} = 8 f^2 \;.
	$$
Thus the typical point-contact conductance decays with distance as a
pure power.  (Note from (\ref{pcontac}) that this is a unique feature
not shared by the average, the variance, or any higher moment.)  In
Ref.~\cite{jmz}, numerical data for $\langle \ln G \rangle$ collected
from the Chalker-Coddington network model, were plotted as a function
of $\ln r$.  The data nicely fell on a straight line, and the best fit
(properly taking into account the statistical errors) was obtained
with $X_{\rm t} = 0.640 \pm 0.009$.  This agrees with the analytical
prediction $X_{\rm t} = 2 / \pi \approx 0.637$, following from $X_{\rm
t} = 8f^2$ and $f^2 = (8\pi\sigma_{xx})^{-1}$ together with the value
$\sigma_{xx} = 1/2$ for the classical conductivity of the network
model.

\section{Discussion}

The present proposal, if correct -- I believe I have given substantial
evidence in its favor and submit it now to the community for judgment
--, opens a new chapter in the theory of the plateau transition of the
integer quantum Hall effect.  For the first time, we have related the
transition to the Lagrangian of a field theory that enjoys manifest
conformal invariance, namely the $A_1|A_1$ nonlinear sigma model with
a WZW term.  This link finally opens the possibility of administering
some of the nonperturbative tools of conformal field theory.
(Although the model is not WZW, it still has a large chiral symmetry
algebra.)  An intriguing aspect of our proposal is that essentially
the same theory appeared in recent work on string propagation in ${\rm
AdS}_3$ backgrounds.  Hopefully, this coincidence will spur
communication between fields and enhance the unity of theoretical
physics.  It also corroborates the status of the quantum Hall effect
as one of the most exciting and profound phenomena brought forth by
recent condensed matter physics.

A remarkable consequence of our proposal is that critical properties
in the QH class seem to be universal only in a restricted sense, as
the true marginality of the coupling $f$ implies the existence of a
line of fixed points, not just an isolated fixed point.  A priori, one
expects that the critical RG trajectories for different members of the
QH class (the network model, for example, the Gaussian white noise
random potential projected on the lowest Landau level, or disordered
Dirac fermions) terminate on different points of the fixed line,
leading to a variable set of exponents.  This seems to be at variance,
at first sight, with what has been observed in numerical and real
experiments.  However, as we have argued, the marginal coupling $f$ is
completely determined by the (renormalized) conductivity governing the
classical motion of electrons near an absorbing boundary. Universality
will therefore persist as long as the classical diffusion over short
distances is governed by a universal conductivity.  It is encouraging
that a number of authors have argued the latter to be true for
incoherent QH systems.

It has been my contention for a number of years that, from a
theoretical perspective, the most naturable probe of critical
transport at the plateau transition is the conductance between two
{\it interior point contacts}.  Such a conductance is represented by a
two-point function of primaries in the conformal field theory.
Moreover, the operators appearing in the field-theoretic
representation of a point contact are operators that create {\it
normalizable} states lying inside the physical Hilbert space.  (In
contrast, the operator for, say, the local density of states creates
unnormalizable states lying outside the physical space.)  The
point-contact conductance therefore is the basic object of the theory,
and any theoretical development is well advised to start from there.
In the present work, we used the short-distance singularity (the
classical limit) of the conductance between two strong contacts to
determine the marginal coupling $f$.  By matching onto the classical
conductivity $\sigma_{xx} = 1/2$ of the Chalker-Coddington network
model at criticality, we found an {\it irrational} coupling $f^2 = 1 /
4\pi$.  We then explained why this value predicts for the typical
point-contact conductance of the network model an exponent $X_{\rm t}
= 2/\pi$, which is confirmed by recent numerical results.

The present work leaves a number of open questions, but provides a
systematic framework for answering them.  The most urgent need is a
calculation of the exponent for the localization length, $\nu$.  The
answer for $\nu$ is not immediate from our results, as this requires
the renormalization of an operator with two derivatives.  (In the
${\rm SU}(2)_1$ WZW model, the most relevant perturbation is ${\rm Tr}
g$.  Such an operator is not available here, as $g$ fails to be
``gauge-invariant''.)  A great many theoretical proposals for this
number have been made in the literature, the most widely cited one
being $\nu = 7/3$ \cite{ms}, and the most recent one $\nu = 20/9$
\cite{leclair}.  From the present theory it seems likely that $\nu$
will be none of these, but some irrational number.


{\bf Acknowledgment}. This paper was conceived and written during a
stay at Princeton University.  I thank P. Sarnak and the Department 
of Mathematics for their hospitality.  I also acknowledge many useful
discussions with D. Bernard in the early part of 1995, when most of 
the structures reported here first became visible.

\end{document}